# From nuclear physics to applications: detectors for beam handling, medical diagnostics and radioactive waste monitoring

Paolo Finocchiaro

INFN Laboratori Nazionali del Sud, via S. Sofia 62, 95123 Catania, Italy

**Abstract**

Nuclear physics experiments are always in need of more and more advanced detection systems. During the last years relevant technological developments have come out with many improvements in terms of performance and compactness of detector materials, transducers, electronics, computing and data transmission. In light of these achievements some applications previously prohibitive, mainly because of size and cost, have become feasible. A few applications of nuclear detection techniques are discussed, starting from the neighboring field of particle beam diagnostics, moving to the medical diagnostics and ending up into the radioactive waste handling. New radiation sensors are shown and explained, as exploited in the DMNR project for the radioactive waste online monitoring which merged into the MICADO Euratom project.

## 1 Introduction

The primary task of experimental physics consists in doing measurements, and this requires to make use of suitable instruments. Modern physics needs to measure quantities not directly accessible to our senses, and for this reason one has to resort to some extension, namely sensors. In particular experimental nuclear physics needs to measure charged particles, neutrons and ionizing electromagnetic radiation, and for this purpose a wide variety of techniques and devices have been developed usually called radiation detectors or simply detectors. The faster and faster technological developments in recent years, mainly aimed at new and better performing nuclear physics measurements, have given rise to many improvements in terms of performance and compactness of detector materials, transducers, electronics, computing and data transmission. Nowadays new high performance materials are available, and previously expensive ones are now easily affordable. Moreover, complex systems are made possible by faster and compact integrated circuits, boards, processors, wired and wireless data transmission interfaces.

In light of these achievements several applications, previously quite difficult or prohibitive, have become feasible as a spillover of nuclear techniques. The following sections will illustrate a few examples of applications of nuclear detection techniques. In section 2 a few methods and devices for the diagnostics of accelerated ion beams are described, all based on photodetection. Section 3 deals with medical imaging and in particular with Positron Emission Tomography (PET). In section 4 new low-cost techniques and detectors for the real time monitoring of radioactive waste are described.

## 2 Applications: beam profiling with scintillators and photodetectors

It is generally acknowledged that an ion beam accelerator is as good as its diagnostic system, and this is why in the particle accelerator community a great care is devoted to the beam diagnostics. Innumerable techniques and devices are feasible, depending on the particle types and energy range and on the particular accelerator features. A comprehensive review of beam diagnostic techniques can be found in ref. [1]. Three practical cases are described in this section, based on scintillators and photodetectors. The sensitivity of particle detection techniques to measure ion beams can be much better than the usual electrical devices (grids, scanning wires), because the electrical sensors are typically sensitive to the overall charge carried by the beam particles, whereas particle detectors respond to the energy of each single particle. Therefore electrical devices are mainly employed when dealing with high intensity beams, whereas in case of low intensity and/or low energy beams, systems based on particle detectors are preferred. Among the particle detectors, scintillators are better suited because of their robustness and ease of use as compared to semiconductors and gas detectors.

### 2.1 A practical case: radioactive ion beam imaging and identification

One of the current hot topics of nuclear physics is the study of nuclei far from stability, i.e. unstable nuclei that undergo radioactive decay toward a more stable configuration. By looking at the nuclei stability chart shown in Figure 1 one can see that the farther a nucleus is from the stability valley (black squares), the faster its decay (shorter half life). Nuclei out of the colored region either have an incredibly short half life or they cannot be formed at all. The properties of nuclei close to the borderline cannot be studied by bombardment with other stable nuclei, as this would imply producing targets of quickly decaying species. The solution is using the radioactive species as a projectile beam onto stable targets. There are basically two methods to produce a radioactive ion beam (RIB): in-flight fragment separation and isotope online separation.

The former (fragmentation ion beam, FRIB) consists in driving a high intensity stable beam onto a suitable thin target, where nuclear fragmentation reactions produce several different nuclear species, usually radioactive and often "exotic" (i.e. very close or just beyond the colored region of Figure 1). By means of magnetic devices one can then select in-flight the species of interest coming out of the thin target, refocusing it into a secondary beam to be sent onto the final target for the experiment. A detailed description of the in-flight technique can be found in the review article [2].

The latter (isotope online, ISOL) still relies on a high intensity stable beam, but in this case it is driven onto a suitable thick target where it is stopped. Several radioactive species thus produced quickly diffuse chemically out of the target where they are ionized. The species of interest is then selected by means of electromagnetic devices and finally reaccelerated by a second accelerator. A detailed description of the ISOL technique can be found in the review article [3].

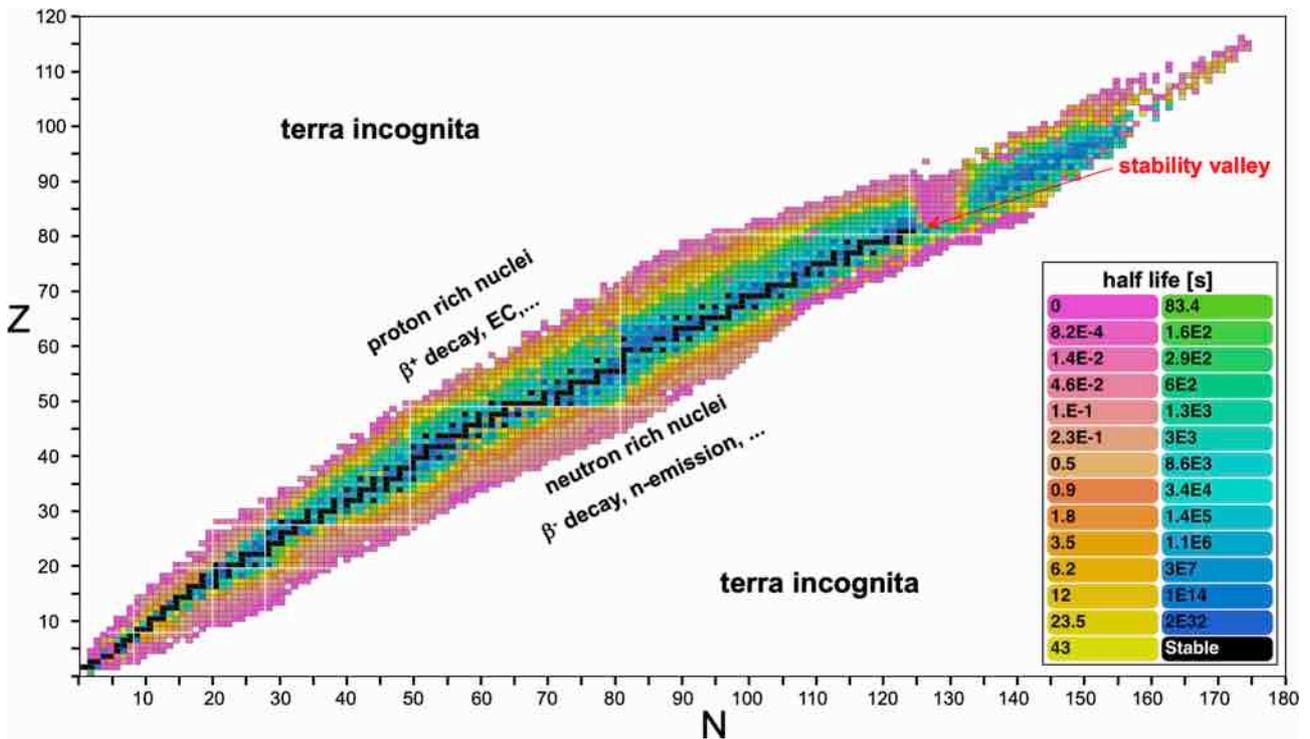

Figure 1. The chart of nuclei, where each combination of neutron and proton numbers (N,Z) corresponds to a small square. The black squares, corresponding to the stable nuclei, lay in the middle of the colored region and represent the so called "valley of nuclear stability". All the other colored squares correspond to unstable nuclear species, with each different color representing a different range of half life.

For ion beam diagnostics one could apparently think of some simple device, but unfortunately this is not the case at all: at different stages and phases of a beam production and acceleration different devices, often complex, are required. In particular during the initial set-up phase the beam can be of low energy, sometimes of low or very low intensity. In some cases it can be both low energy and low intensity, not to mention the case when it is also radioactive with an intensity range starting from few particles per second.

There are many possible sensors and techniques for beam detection: direct electromagnetic (scanning wires, faraday cups), optical (scintillators and scintillating screens), semiconductors, diamond detectors, gas detectors, secondary electrons emission, Cherenkov, and others. The focus of this section will be on scintillators and scintillating screens.

A scintillating screen is a plate made of a scintillating material, i.e. a material that emits light when hit by ionizing radiation (gamma or charged particles) [4]. Basically there are two classes of scintillator detectors: organic and inorganic. Organic scintillators are plastic and are easily damaged by a charged particle beam. Inorganic scintillators, typically crystals, are more robust with typically higher light yield, and therefore are usually preferred for beam diagnostics purposes. A wide variety of scintillating materials is available, with different features in terms of light yield, emission spectrum, time duration of the light pulse. The task performed by a scintillating screen is to produce in real time a visible image of the beam projection in a plane perpendicular to the beam direction, therefore providing information on the beam intensity, shape, size and position, useful for its tuning and alignment. A screen, installed on an in-beam/off-beam actuator, is usually oriented at 45° with respect to the impinging beam direction, so that it can be viewed through a glass porthole by means of a camera outside the beam line at 90°. Figure 2 shows a simple setup used to test a scintillating screen by means of a radioactive source. In this case a $^{90}$Sr beta source was employed, which emits high energy electrons up to 2280 keV. A metal collimator in front of the source restricted the angular range of the electrons, in order to simulate a beam. The scintillator plate material was CsI(Tl) which has a light yield of about 50000 visible photons per MeV of deposited energy [5].

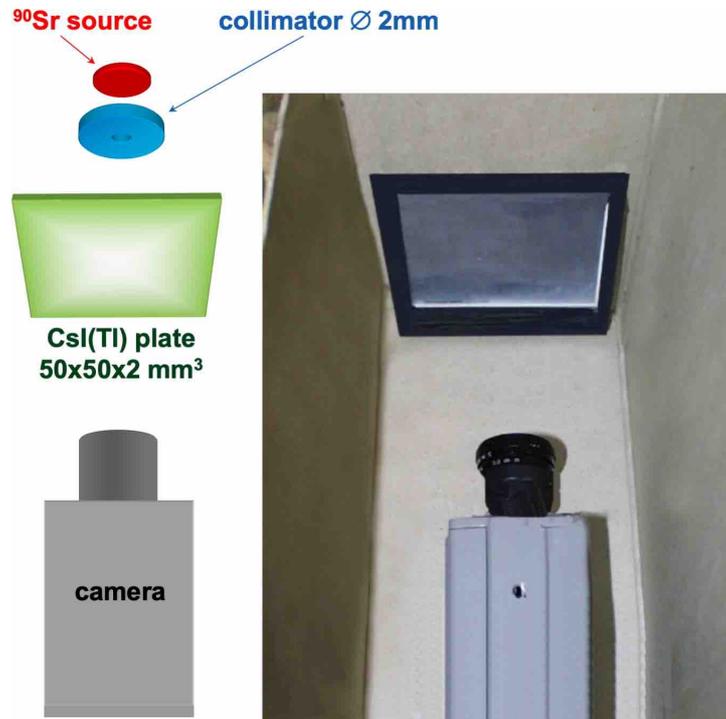

Figure 2. Simple setup used to test a scintillating screen by means of a radioactive source. A $^{90}$Sr beta source was employed, which emits high energy electrons up to 2280 keV. A metal collimator in front of the source restricted the angular range of the electrons, in order to simulate a beam. The scintillator plate material was CsI(Tl) which has a light yield of about 50000 visible photons per MeV of deposited energy.

Figure 3 shows the result of a test with the previous setup. The top-left panel contains the 3D representation of the raw data, affected by the systematic noise of the CCD camera. In the top-right panel the systematic noise pattern of the CCD camera is shown, that was obtained by closing the camera shutter. The noise-subtracted image is then reported in the bottom-left panel, and is also shown as realistic 2D beta "beam" image in the bottom-right panel. The test proved that in addition to imaging an ion beam such a scintillating screen could be used to produce an image of a beta radioactive source. This paved the way to an interesting application, namely imaging a low energy and low intensity unstable $^{8}$Li beam and its stable isotopic counterpart $^{7}$Li, produced by the ISOL method, by means of a device called LEBI (Low Energy Beam Imager) [6].

LEBI, shown in Figure 4, makes use of two scintillators: a CsI(Tl) plate and a bulk fast plastic BC408 [7] scintillator. The arm holding the device can be remotely moved in order to place it off-beam or to have the beam hitting in the position *b*, *c* or *d*. In order to start the beam tuning one had to first extract and transport an "easy" stable nuclear species, as close as possible to the unstable one, by tuning the magnetic selector after the thick target diffusion to accept the stable $^{7}$Li$^{1+}$. By choosing the *c* LEBI position the stable low energy beam (10 keV kinetic energy) impinged directly onto the plate, thus producing a visible image of the beam profile that was acquired by means of a CCD camera placed off-beam. The optimum beam shape required for the further transport was a thin vertical structure (Figure 5 left, despite the presence of a beam halo to the right of the main spot).

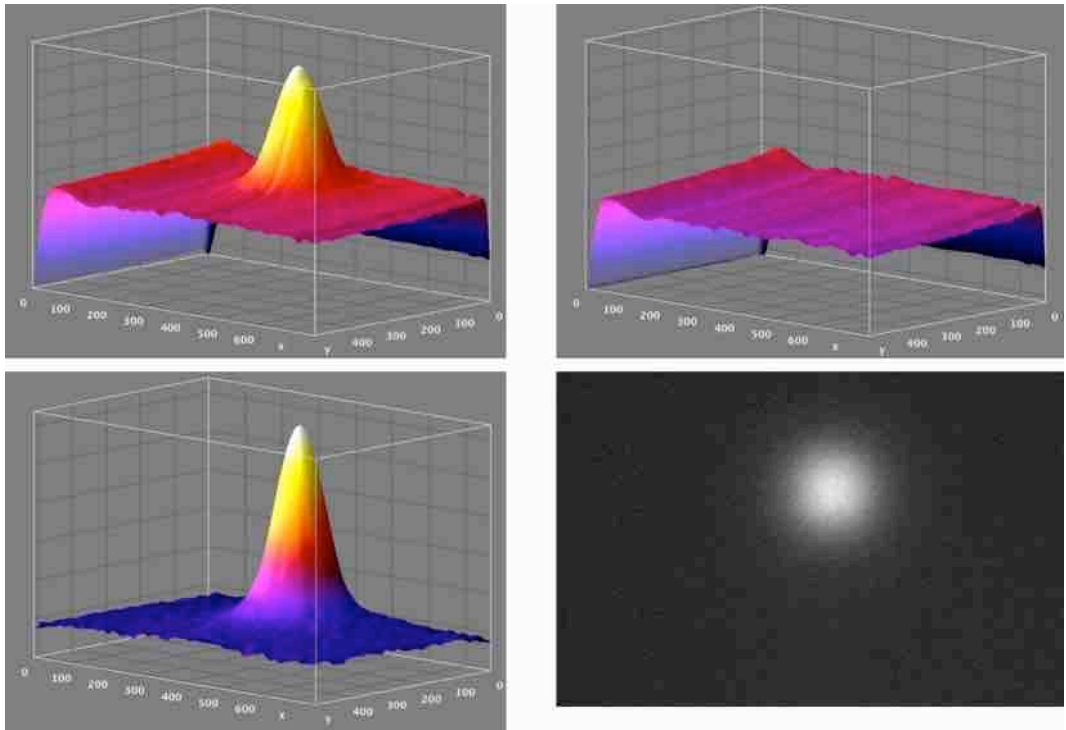

Figure 3. Result of a test with the setup of Figure 2. Top-left: 3D representation of the raw data, affected by the systematic noise of the CCD camera. Top-right: the systematic noise pattern of the CCD camera. Bottom-left: noise-subtracted image. Bottom-right: final 2D beta "beam" image.

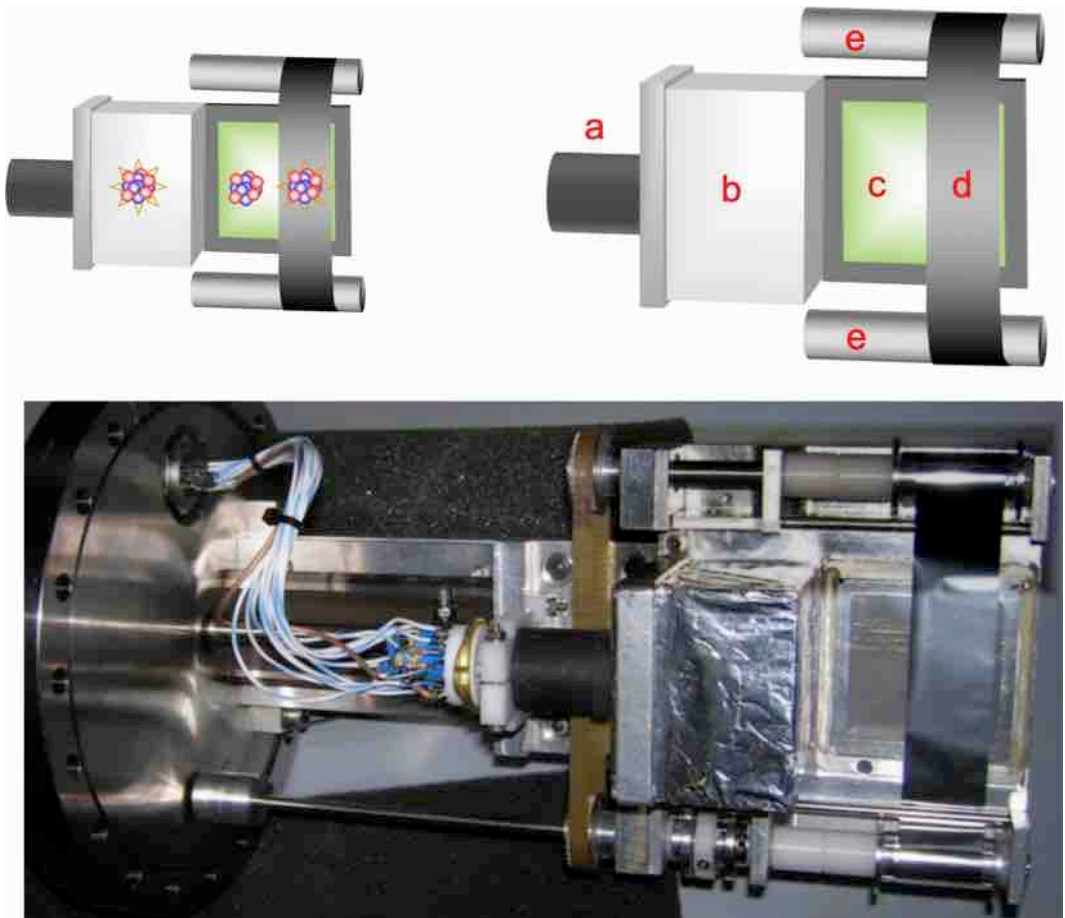

Figure 4. The LEBI device. (a) Photomultiplier tube. (b) Plastic scintillator BC408. (c) CsI(Tl) scintillating plate. (d) Mylar tape to implant the radioactive beam. (e) Mechanical spools for the mylar tape. See the text for details.

The LEBI position was then switched to *d*, and the field of the magnetic selector was suitably scaled to accept the unstable $^8$Li$^{1+}$. In this configuration the low energy beam was implanted onto the mylar tape in contact with the CsI(Tl) plate. $^8$Li is a beta decaying nucleus, with a maximum energy of the emitted electron up to 12.9 MeV. The electrons emitted in the forward direction easily crossed the thin mylar tape and reached into the scintillating plate, where they produced a visible image of the beam profile acquired by means of the CCD camera and shown in Figure 5 right. The half life of $^8$Li is 840 ms, therefore after stopping the beam the scintillation light produced by the beta decays fell down in few seconds. After 10-20 seconds the plate was totally dark again, but this would not be the case with a longer lived unstable beam. This is the reason of the mylar tape, as in such a case instead of contaminating the plate by implanting the unstable nuclei on it, they would be implanted on the tape. For each beam measurement one should roll the tape up to a fresh position.

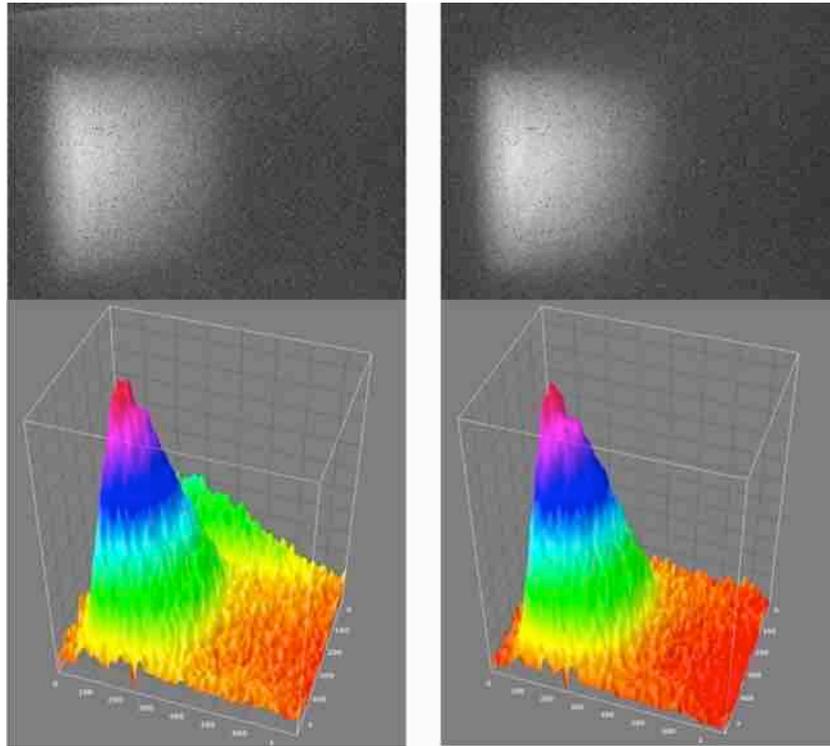

Figure 5. Left: the stable $^7$Li beam profile. Right: the radioactive $^8$Li beam profile. See the text for details.

A final check that the selected unstable beam species was $^8$Li, and not for instance the isobar $^8$B (half life $T_{1/2}$=770 ms), was done by means of the plastic scintillator read out by the photomultiplier tube, respectively *b* and *a* in Figure 4. Such a fast scintillator made possible to count the beta decay events from the tape, and the results are reported in Figure 6 as measured counting rates as a function of time. The plot, from left to right, starts with the beam already on. Then a sequence of beam off, on and off again shows the expected exponential behavior as the tape (de)populates with beam particles. A fit with the proper populating and decaying exponential functions gave 840±26 ms for the half life $T_{1/2}$ that is exactly the half life of $^8$Li.

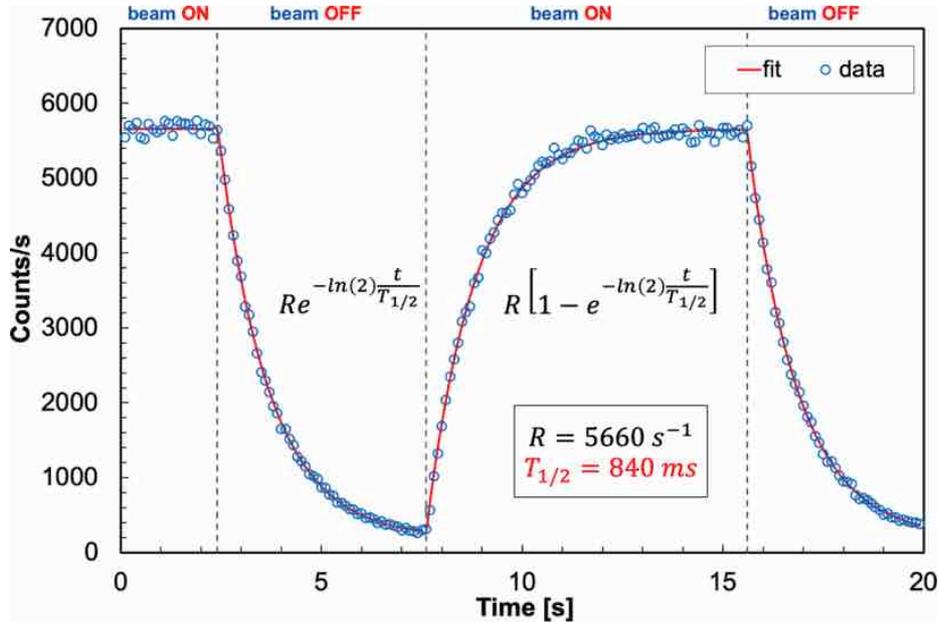

Figure 6. Counting rate as a function of time measured in intervals of 0.1 s by means of the plastic scintillator and the photomultiplier tube (*b* and *a* in Figure 4). From left to right a sequence of beam on, off, on, off shows the expected exponential behavior as the tape (de)populates with beam particles. A fit with the proper populating and decaying exponential functions gave 840 ms for the half life $T_{1/2}$ that is exactly the half life of $^8$Li.

*2.2 A practical case: microbeam diagnostics*

Another interesting tool for several applications is represented by ion beams with microscopic spot size. As example applications it is worth mentioning the biological study of microdosimetry on single cells [8] and the production of micro-optical components by means of controlled damage on samples followed by chemical treatments, technique also named Deep Lithography with Ions (DLI) [9]-[12]. There are two methods to produce microbeams: focusing, by means of electromagnetic lenses, and collimation. A focused microbeam requires the sample under irradiation to be placed in the focal plane, where the beam has the minimum spot size. Conversely, a collimated microbeam gives more freedom on the position of the sample along the beam direction. In both cases a very low beam intensity is usually employed, sometimes even down to single particles. This implies that the beam diagnostics plays a relevant role in the handling of such beams. Several dedicated diagnostic tools have been developed so far and are currently in use, here I describe the features and use of a particular scintillating plate called SFOP (Scintillating Fiber Optic Plate).

A SFOP is obtained by cutting a slice from a bunch of optical fibers, made from scintillating material, glued together [13]. The plate thus obtained has a honeycomb-like structure, that produces scintillation light when hit by ionizing radiation. The light is emitted isotropically and mostly stopped by suitably interspersed black fibers called Extra Mural Absorbers (EMA), but a fraction is trapped along the fibers and transported to the back face where it can be watched by a camera (Figure 7). The advantage of using a SFOP instead of a normal plate as in section 2.1 is in the spatial resolution. In a normal plate the scintillation image profile created by the beam from the front face is slightly blurred when looked at from the back, due to thickness of the plate itself and to the production of the light at varying depth, with a typical resolution of the order of 1 mm. Conversely, in a SFOP the spatial resolution is given by the fiber diameter that can be as low as 10 μm. Figure 7 illustrates a typical setup employed for the production of micro-optical components. A proton beam of 9 MeV energy was first sent through a 1 mm diameter and 12 cm long pre-collimator to reduce the beam size and divergence. A micro-collimator, made from copper 300 μm thick, hosted a set of holes from 1 mm to 20 μm diameter and could be vertically positioned at will by means of a remotely controlled electromechanical actuator. Following the two collimators there was a sample, made from PMMA (Poly-Methyl-Meta-Acrylate, i.e. plexiglass), on which a controlled damage was produced by moving the sample along a microscopic pattern at a predefined speed. Before and after the irradiation the sample was moved off-beam, and the beam profile was checked by means of a SFOP followed by a mirror at 45° and a videocamera. After the irradiation a suitable chemical treatment would etch away the irradiated parts thus realizing microscopic lightguides, fiber holders, prisms, and/or a different treatment would make pointlike irradiations to swell with spherical curvature thus obtaining micro-lenses (Figure 8).

The beam diagnostics in this application is crucial, because one has to ascertain the correct alignment that gives rise to the required beam size and shape. As for the beam intensity, a prior calibration procedure of the current-to-light yield makes possible to select the desired beam current and sample speed. Figure 9 shows a micro-beam profile image and its zoom in false colors, the FWHM width was ≈ 80 μm.

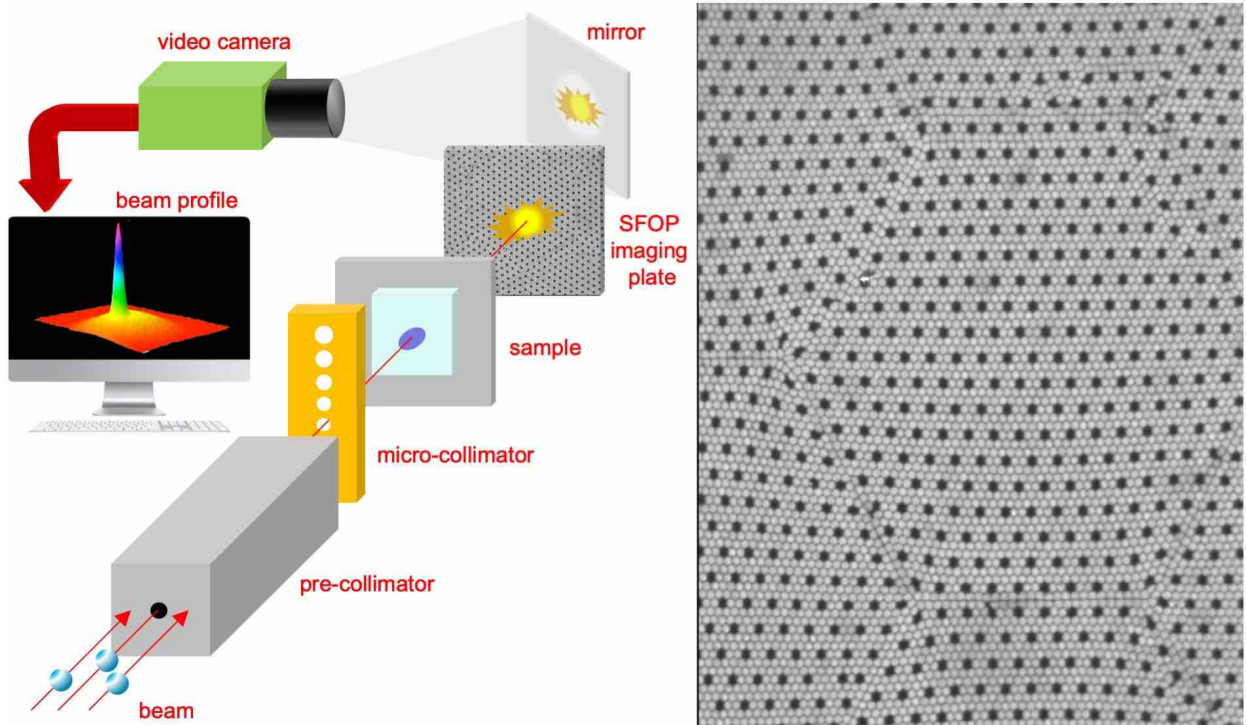

Figure 7. Typical setup for the production of micro-optical components. A proton beam of 9 MeV energy was first sent through a 1 mm diameter and 12 cm long pre-collimator. A micro-collimator with a set of holes from 1 mm to 20 μm diameter could be vertically positioned by means of a remotely controlled actuator. Following the two collimators there was a sample, made from PMMA, on which a controlled damage was produced by moving the sample along a microscopic pattern. The beam profile was checked by means of a SFOP followed by a mirror at 45° and a videocamera.

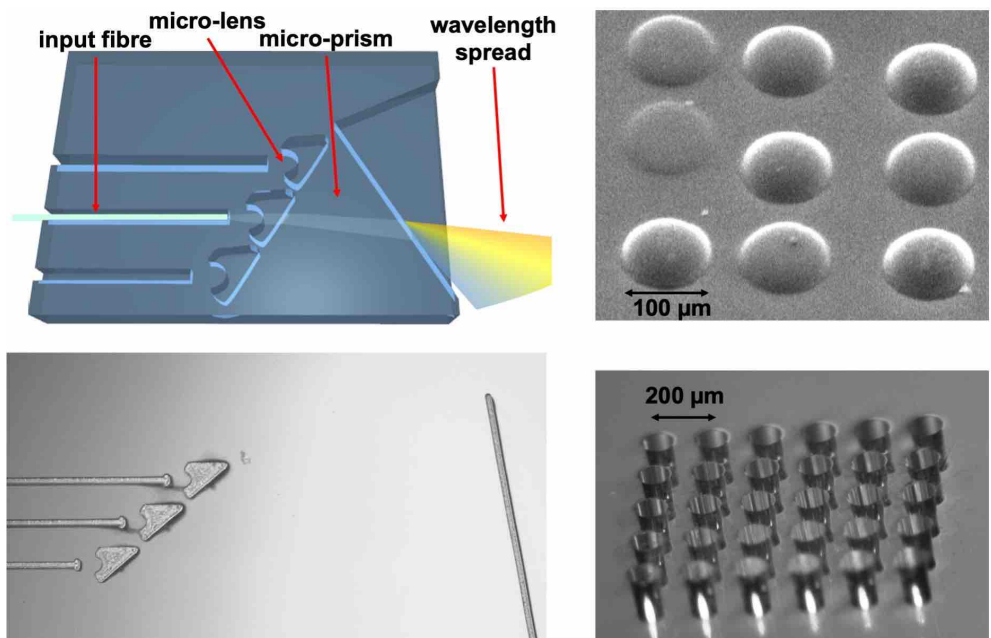

Figure 8. Examples of micro-optical components produced by means of the DLI technique. Left: microprisms with fiber holders and cylindrical lenses. Top right: micro-lenses. Bottom right: array of conical micro-holes for multifiber connection.

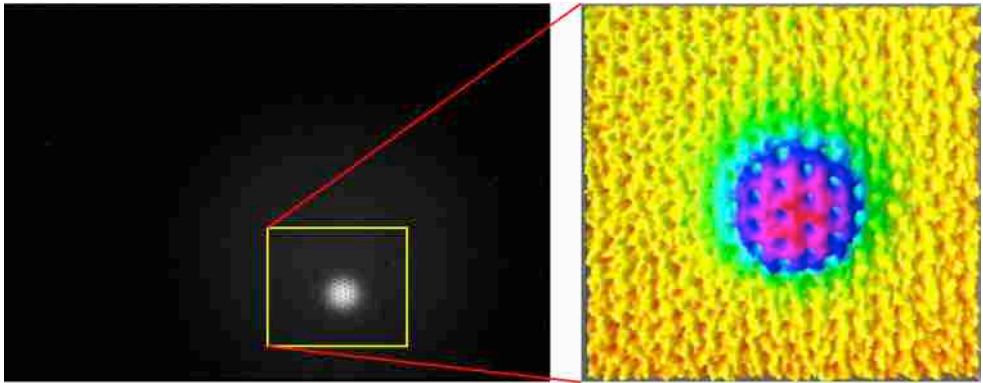

Figure 9. A micro-beam profile and its zoom in false colors, the FWHM width was ≈ 80 μm.

## 2.3 A practical case: scintillating fibers

The scintillating fibers can also be used individually for ion beam diagnostics, indeed the scintillation light produced upon impact of the beam particles can be collected by a photomultiplier tube (PMT) coupled to one fiber end. Scanning the beam by moving the fiber across it, one can reconstruct the beam intensity as a function of one coordinate. If using two mutually perpendicular fibers one can thus reconstruct the X and Y beam intensity profiles. Employing plastic scintillating fibers [14], which feature very fast light pulses of the order of few ns, makes it possible to count beam particles one by one up to several million counts/s. When the counting rate becomes too high, at each fiber position one can simply measure the current level of the PMT. Both operations are possible simultaneously by means of a suitable current-to-voltage circuit with two independent outputs: a DC voltage proportional to the PMT output current (high rate condition) and a pulse signal (lower rate condition). In Figure 10 a device called Fiber Based Beam Sensor (FIBBS) is shown. It has two scintillating fibers, mutually perpendicular, installed on a single compact PMT. The fibers and PMT assembly can be moved forth and back in front of the ion beam by means of a stepping motor, with the duration of each step selectable by software in order to collect enough counting statistics in each position. A suitable shield prevents the possibility that the beam hits both fibers at once. The FIBBS reconstructs the beam intensity profile along two directions orthogonal to each other, with a typical duration of about one second for a full beam scan. In case of very high beam intensity the plastic fibers, which are not highly resistant to radiation, can be replaced with glass scintillating fibers [13] which are more radiation hard. The glass fibers have milliseconds decay time for the scintillation light, so they are only used in the DC mode.

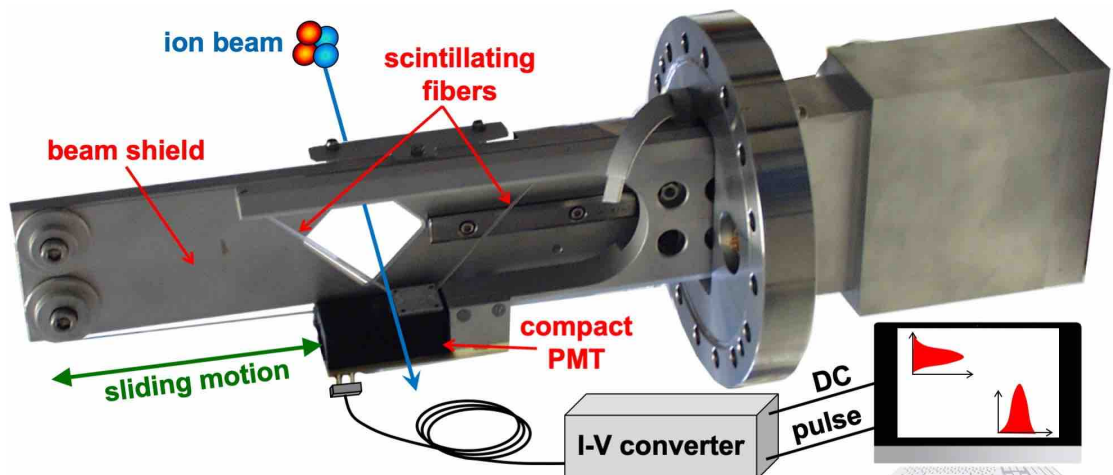

Figure 10. Fiber Based Beam Sensor (FIBBS). Two scintillating fibers, mutually perpendicular and installed on a single compact PMT, can be moved forth and back in front of the ion beam by means of a stepping motor in about one second. The FIBBS reconstructs the beam intensity profile along two directions orthogonal to each other.

## 3   Applications: medical imaging with (nuclear) physics techniques

The medical imaging science started out in the early 1900 following the discovery of X-rays by Wilhelm Röntgen. Radiography exploits the absorption of the ionizing X radiation when crossing the human body: the denser the tissues and the higher its average atomic number <Z>, the stronger the absorption. Therefore in a radiograph the black color indicates radiation that did not interact with the body, whereas the white color indicates total absorption. All of the gray shades correspond to different levels of absorption, i.e. density and thickness.

At the beginning the image was directly impressed on the photographic film by the radiation, and this implied high radiation doses released to the patients. Then photosensitive films with a high-yield scintillating emulsion layer were developed, whereby the radiation produces scintillation light that is recorded on the film, thus resulting in much lower exposure required and consequently much lower doses to the patient. The typical image produced by X-rays is a plane projection from a single direction. Figure 11 shows an example of cervical spine radiograph.

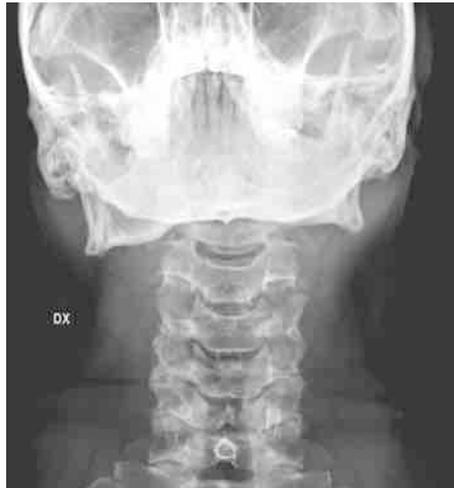

Figure 11. Example of a cervical spine radiograph.

In 1969 Godfrey Hounsfield (engineer) and James Ambrose (neuroradiologist) invented the Computed Tomography (CT) scanner, and in 1974 they jointly won the BJR Barclay prize. In 1979 Hounsfield was awarded the Nobel Prize in Physiology and Medicine. The basic principle of operation of a CT scanner is to take a set of X-ray images from a large number of different directions, and then to apply special algorithms to perform the so-called back-projection in order to obtain a 3D representation of the density profile of the object under observation. In practice a rotating head is employed, which hosts an X-ray tube and a set of opposite small detectors. While rotating around the patient, intensity data measured by the detectors are recorded numerically by a computer system. This process is repeated many times by varying the longitudinal position of the patient with respect to the rotating head, thus resulting in a collection of numerical data from a set of adjacent slices. The final result of the back-projection is a 3D density map of the organs under observation, that can be examined on a screen and/or printed in form of slices and also be displayed and/or printed as 3D views (Figure 12). An important variation of this technique consists in administering to the patient a special drug, called contrast medium, containing a high Z element and capable of concentrating into particular organs or tissues.

A different imaging concept is exploited with the Single Photon Emission Computed Tomography (SPECT). It consists in administering to the patient a drug containing a short-lived gamma-emitting radioisotope, typically $^{99}Tc^m$, a metastable isotope of Tecnetium with 6 hours half life, and then detecting the gamma rays mainly emitted from the region where the absorbing organ is located. In order to produce images one has to perform the back-projection, and this requires the acquisition of numerical data in form of gamma-ray counting rates. Unfortunately, at variance with the X-rays which one can easily produce with a well defined directionality, gamma rays are emitted isotropically and from inside the patient. Therefore, in order to produce correct projections, one has to select those coming from a given direction. This is achieved by means of gamma cameras consisting in a set of collimators, made from a thick high-Z material, installed in front of a position sensitive gamma ray detector. Figure 13 illustrates the SPECT technique, with three identical rotating gamma cameras in order to speed up the data collection and to optimize the dose to the patient by minimizing the amount of the administered drug. The SPECT produces low resolution images/slices, but it can be highly tissue-selective. Therefore a tracer drug mainly accumulating in cancer tissue will produce a high contrast image of the tumor. By superimposing such an image with a standard X-ray o CT scan one can easily spot and locate tumor masses and metastases. Interesting improvements of SPECT spatial resolution can be obtained by means of more sophisticated collimators [15].

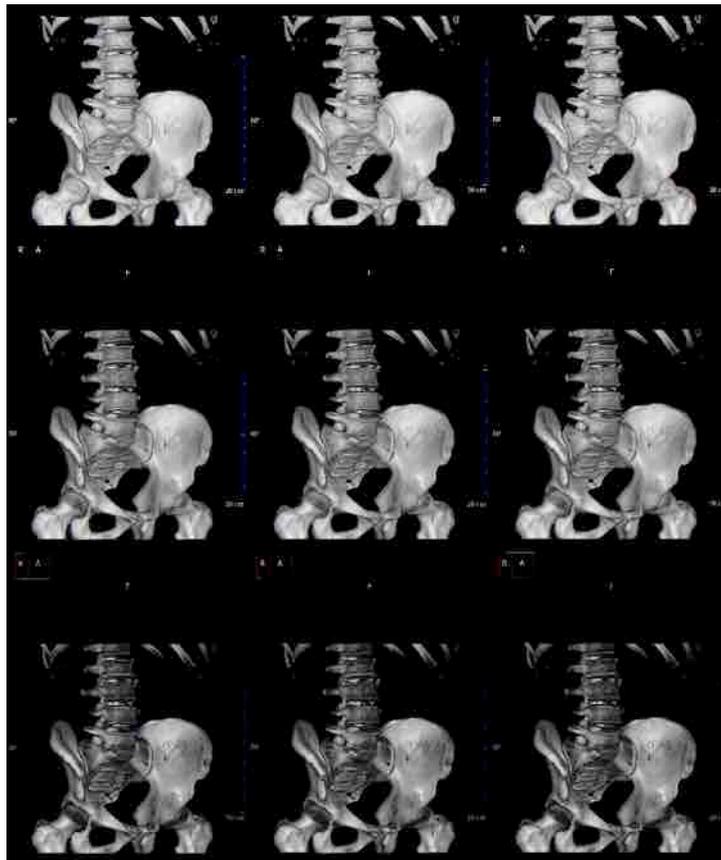

Figure 12. Example of a 3D view of a lumbosacral spine and pelvis CT scan under nine different contrast options to increase the visual dinamic range and facilitate a detailed examination.

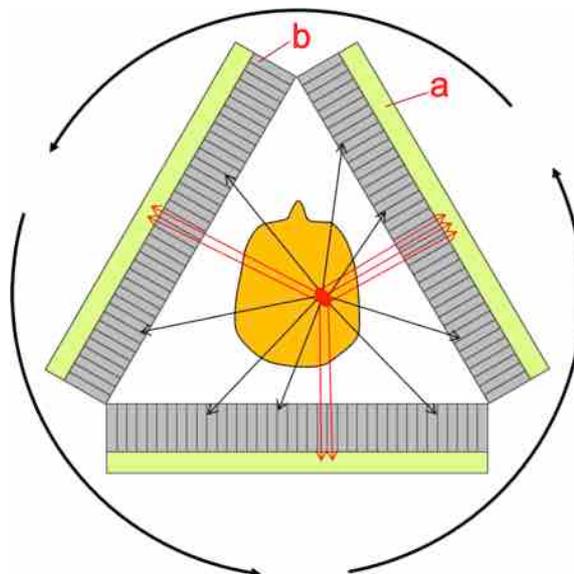

Figure 13. Sketch of the SPECT technique, with three identical rotating gamma cameras in order to speed up the data collection and to optimize the dose to the patient by minimizing the amount of the administered drug. (a) The position sensitive gamma ray detector. (b) The set of thick high-Z collimators.

## 3.1 Positron Emission Tomography (PET)

An intrinsic limitation to the SPECT imaging comes from the counting statistics: in order to build the images one has to select the incoming direction of the gamma radiation, thus losing most of it in the collimators.

The Positron Emission Tomography (PET) overcomes this limitation by making use of beta+ radioactive tracers, in most cases Fluorodeoxyglucose enriched in $^{18}$F (18-FDG). The positron (positive electron) emitted in the beta+ decay, with maximum energy 250 keV, stops almost immediately in matter where it annihilates with one electron giving rise to two gamma rays of 511 keV each, emitted back to back. By detecting both gamma rays the emission direction is determined, and a collection of such events can be back-projected, with algorithms similar to the CT scan reconstruction, and produce a 3D map of the 18-FDG distribution inside the patient, that can be examined on a screen and/or printed in form of slices and also be displayed and/or printed as 3D views. Similarly to SPECT, PET is highly tissue-selective or better organ-selective: FDG is a glucose and its uptake by tissues is a marker for the tissue uptake of glucose, which in turn is closely correlated with certain types of tissue metabolism, and is retained by tissues with high metabolic activity, such as most types of malignant tumors [16],[17].

An overall sketch of the operational principle of PET is depicted in Figure 14. After being administered the 18-FDG tracer, the patient is positioned inside a ring-shaped detector array. The data are acquired whenever the system detects a coincidence between two detectors located at large mutual angles (around 180°). An additional constraint is that both the measured gamma rays must have 511 keV energy (Figure 15), useful to reject most of the spurious coincidences due to Compton-scattered gamma rays. For all events fulfilling these two conditions, two values $r$ and $\theta$ are computed and are used as coordinates to build a 2D histogram usually called sinogram because of its overall sinusoidal shape, as shown in Figure 16. $r$ is the distance of the Line Of Response (LOR) from the ring center, $\theta$ is the angle between the horizon and the LOR. The recorded sinogram is later fed into a back-projection algorithm to produce a 3D representation and slices. The spatial resolution of the final images is dominated by the individual detector size. However, the smaller the detector the lower its solid angle and consequently the counting statistics, implying that in order to have an equivalent statistics one should use a larger quantity of tracer i.e. dose released to the patient. A tradeoff between typical dose, detector size and cost is needed. The gamma ray detectors employed in the first PETs were inorganic scintillator crystals made from Gadolinium Oxyorthosilicate (better known as GSO) read out by means of PMTs [18]. Figure 17 shortly summarizes the features of the main medical imaging techniques based on ionizing radiation.

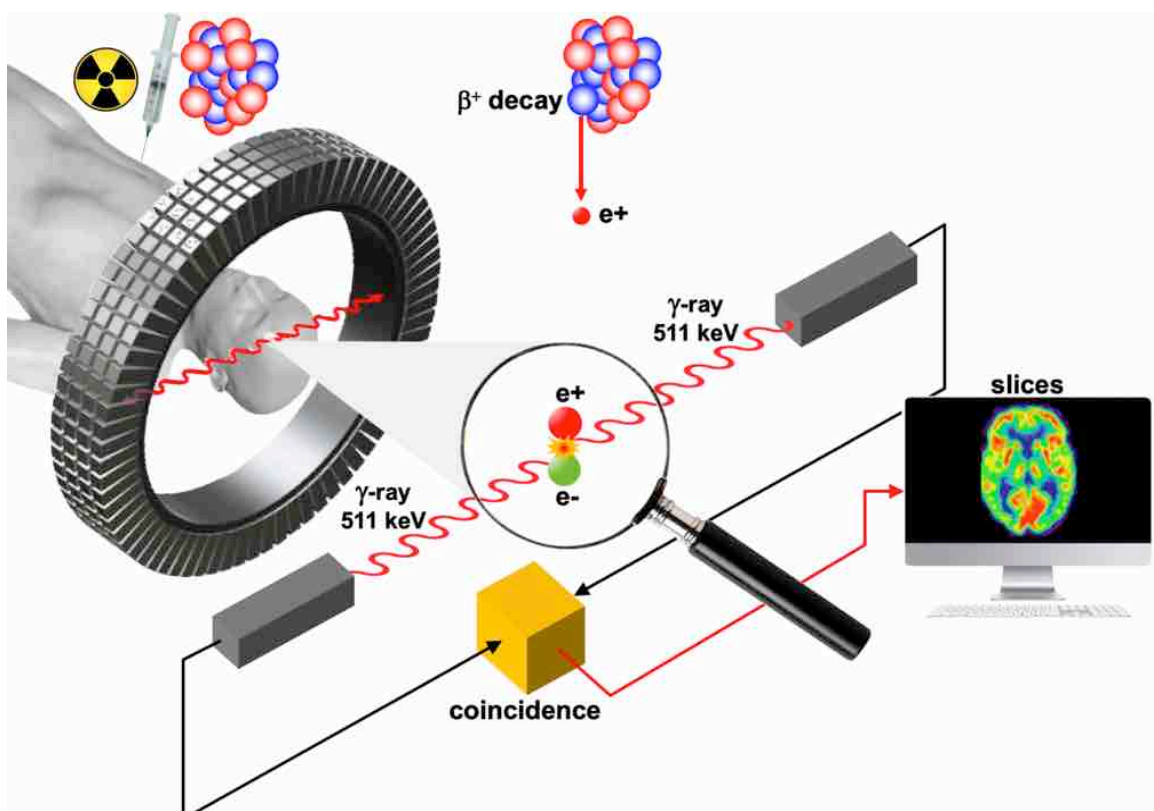

Figure 14. Sketch of the operational principle of PET. After being administered the beta+ radiotracer, the patient is positioned inside a ring-shaped detector array. The data are acquired whenever the system detects a coincidence between two detectors located at large mutual angles

(around 180°). An additional constraint is that both the measured gamma rays must have 511 keV energy.

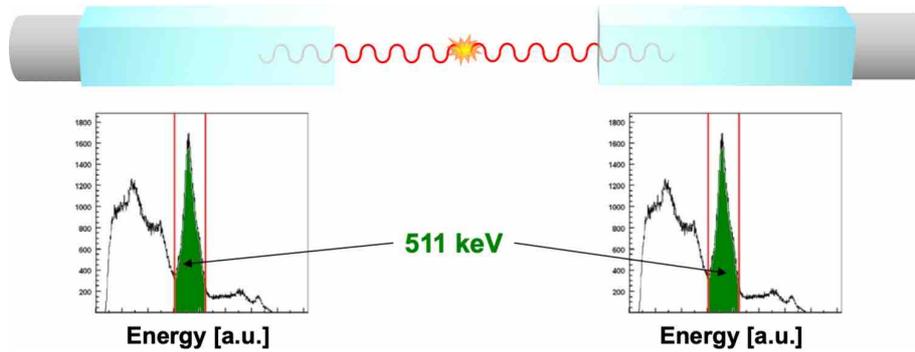

Figure 15. Energy constraint for the PET data. In addition to detect a coincidence between two opposite detectors, the measured energy of both gamma rays must be 511 keV.

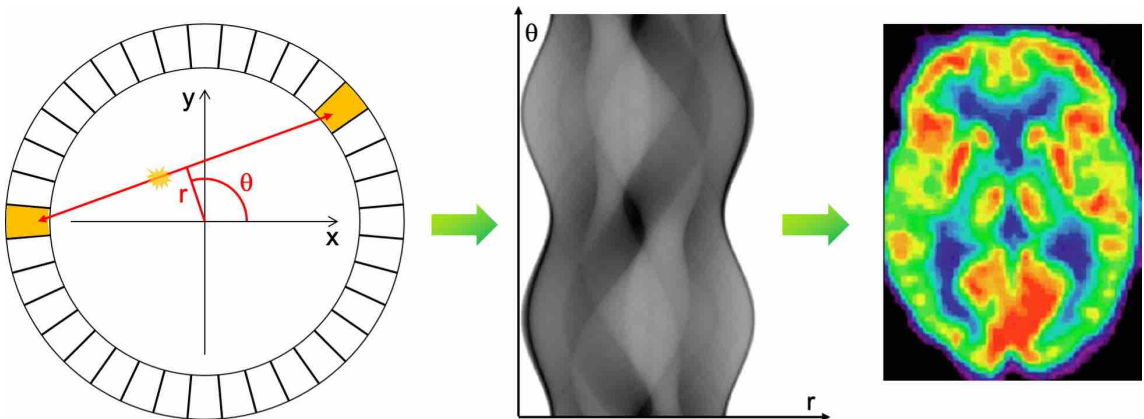

Figure 16. For each event fulfilling the coincidence and energy constraints two values *r* and *θ* are computed (left) and used as coordinates to build a 2D histogram called sinogram (middle). A back-projection algorithm transforms the sinogram into a 3D representation and slices (right).

| | probe | tool | info about... | image type |
|---|---|---|---|---|
| | X-rays | Radiography | anatomy (tissue density) | projection |
| | X-rays | Computed Tomography (CT) | anatomy (tissue density) | 3D and slices |
| | γ-rays | SPECT | physiology (cells/organs in operation) | 3D and slices |
| | γ-rays | PET | physiology (cells/organs in operation) | 3D and slices |

Figure 17. Summary of the features of the main medical imaging techniques based on ionizing radiation.

### 3.2 *Time Of Flight Positron Emission Tomography (TOF-PET)*

An interesting and very promising improvement of PET is represented by the Time Of Flight PET (TOF-PET) [19],[20]. The first suggestions were made in the 70s, and the first practical tests in the 80s. However, the real breakthrough came in recent years with the development of new inorganic scintillators, like Lutetium Oxyorthosilicate (LSO) or Lutetium Yttrium Oxyorthosilicate (LYSO) [5],[7],[18], and at the same time of new compact solid state photomultipliers (SiPM) [21]-[32]. In addition to the measurement of LOR and gamma rays energy, TOF-PET also measures the time difference between the detection of the two gamma rays. The new scintillators have high effective atomic number Z, and a better light yield that provides a better energy resolution. This, along with a light decay time of ≈40 ns, makes possible to attain an unprecedented time resolution with this kind of detectors. The SiPM is a rather new photodetector featuring, among other properties, very good photodetection efficiency, intrinsic signal amplification of the order of $10^6$, low-voltage operation (30-70 V), good energy resolution, superior timing properties. Its main disadvantage, namely the few millimeter small size, turns into an advantage for TOF-PET as it makes possible to set up arrays of high performance tiny detectors [33],[34],[35].

In TOF-PET sinograms are not used, data are collected in list mode event by event: for each event recognized as a two-detector coincidence, the identifiers of the two detectors hit are stored along with the two measured energy and time-of-flight values (Figure 18). These data can be replayed later with a suitable back-projection algorithm to calculate the LOR (Figure 16 left), to refine the energy window filter (Figure 15), and to apply a time filter as shown in Figure 19. Indeeed, with the conventional PET the beta+ decay, origin of the two gamma rays, could have occurred at any position on the LOR. Conversely, in TOF-PET the difference between the two arrival times makes it possible to determine the decay position along the LOR, calculated as

$$x = c \frac{t_j - t_i}{2} \qquad (1)$$

where $x$ is the coordinate on the LOR with respect to its midpoint, $c$ is the speed of light, $t_j$ and $t_i$ are the measured times. Assuming the same time resolution $\Delta t$ for the two detectors the uncertainty is given by

$$\Delta x = c \frac{\Delta t}{\sqrt{2}} \qquad (2)$$

The sketch in Figure 19 right illustrates the improvement introduced by TOF-PET: event by event, in addition to the direction of the LOR, the back-projection algorithm can rely on the position of the emission point along the LOR, provided that the time resolution is good enough to significantly restrict the emission region. The time resolution achievable with the currently available detectors (LYSO+SiPM) is of the order of 250 ps, corresponding to a spatial resolution of *$\Delta x \approx 5$ cm*. This information introduces a statistical weight on the emission point, and therefore improves the definition of the final images. Moreover, it also helps suppressing the spurious noise (artifacts) caused by non-physical events corresponding to positions outside the detector ring [19]. A very promising application of the TOF-PET with miniature detectors is for the prostate tumor diagnosis and follow-up, by means of an endorectal probe plus a plane detector to be positioned over the patient's abdomen, at variance with the ring geometry, as employed in the TOPEM project [20].

Any spatial resolution improvement must also take into account the blurring caused by the parallax error, since the two-gamma emission can occur from an off-center extended source. In reality the LOR is not a line but rather a cylinder or a prism, due to the finite size of the detectors. Figure 20 explains why it is important to correct the parallax effect by determining event by event the Depth Of Interaction (DOI) in the detectors, in order to make the most of their spatial resolution. An example of miniature detectors for TOF-PET [33],[34],[35] is based on 1.5 mm × 1.5 mm × 10 mm LYSO scintillators, each one read out by means of two SiPMs (Figure 21). In such a configuration, if a gamma-gamma coincidence is detected for instance in detectors #k and #m, the eight quantities *($E_f^k$,$t_f^k$), ($E_b^k$,$t_b^k$), ($E_f^m$,$t_f^m$), ($E_b^m$,$t_b^m$)* are measured, with the lower index *f* standing for *front*, *b* standing for *back*, and the higher indexes *k* and *m* referring to the detector number. Therefore $E_f^k$ indicates the scintillation light signal detected by the front SiPM of detector number *k*, $t_f^k$ will be the time measured by the same SiPM, and so on for the other values. The front and back time values should be basically the same, as the difference of the order of 10 ps due to the 10 mm crystal length cannot be appreciated. One can choose to only use one of the two values, or better take the average to decrease the uncertainty by a factor square root of 2.

The scintillating crystals are typically wrapped with high quality optical reflectors, in order to prevent the light from being attenuated before reaching the photodetector. Indeed the attenuation of light travelling in a linear scintillator follows a decreasing exponential law governed by a constant called attenuation length. The larger the attenuation length, the higher the amount of light reaching the photodetector. In this particular case, however, it can be shown that by wrapping the crystals with a non-optimum reflector material, so to have an attenuation length of the same order of the crystal length, simple mathematical calculations make it possible to determine the DOI and gamma energy [35]. Indeed one has:

$$DOI \propto \ln\frac{E_f}{E_b} \quad (3)$$

$$E_\gamma \propto \sqrt{E_f \cdot E_b} \quad (4)$$

Such a configuration, as tested on the bench, provides a time-of-flight resolution $\Delta t \approx 200\ ps$ corresponding to a position resolution along the LOR $\Delta x \approx 4\ cm$, energy resolution of the 511 keV gamma rays $\Delta E \approx 75\ keV$, DOI resolution $\Delta_{DOI} \leq 1\ mm$. A better timing resolution could be obtained by using shorter crystals with single end readout, but at the expense of the detection efficiency. Therefore the TOPEM concept represents a tradeoff between detection efficiency and timing.

Figure 18. Scheme of the list mode data acquired in TOF-PET.

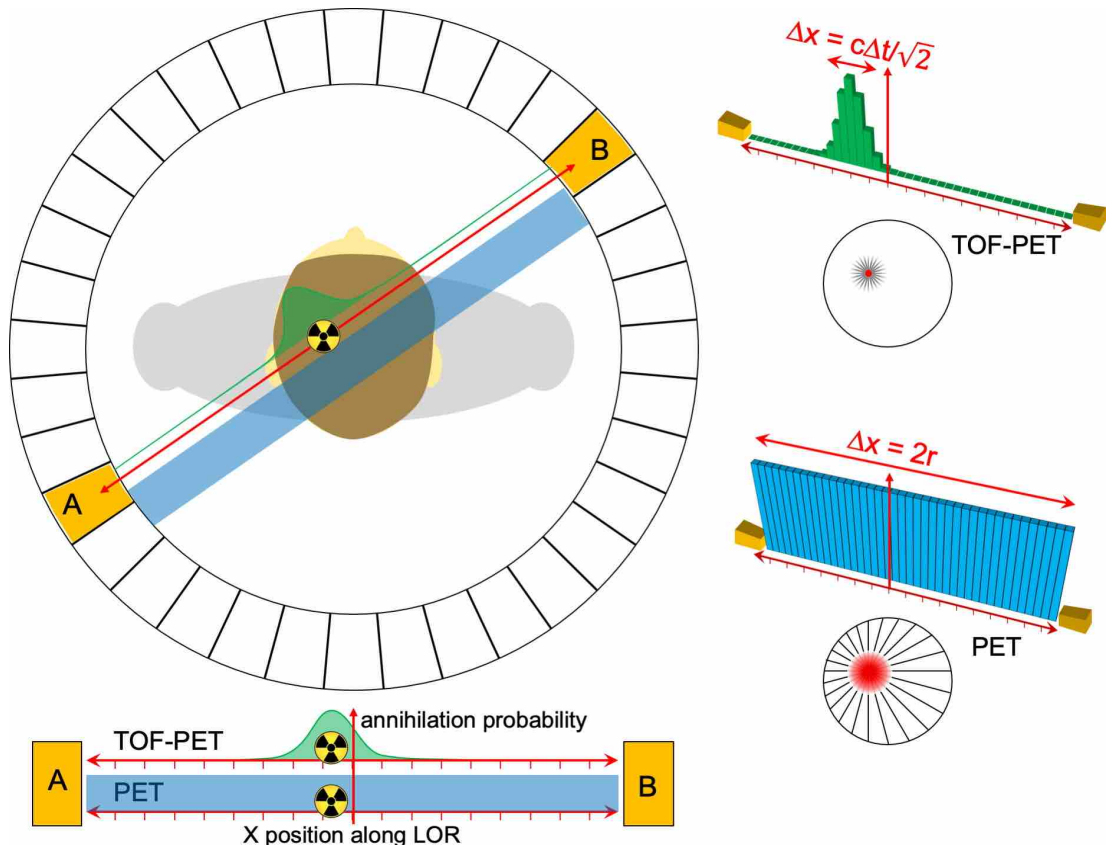

Figure 19. Improvement introduced by TOF-PET. In addition to the direction of the LOR, Eq.1 and 2 provide a peaked probability distribution for the emission point position, that in the case of simple PET is uniform.

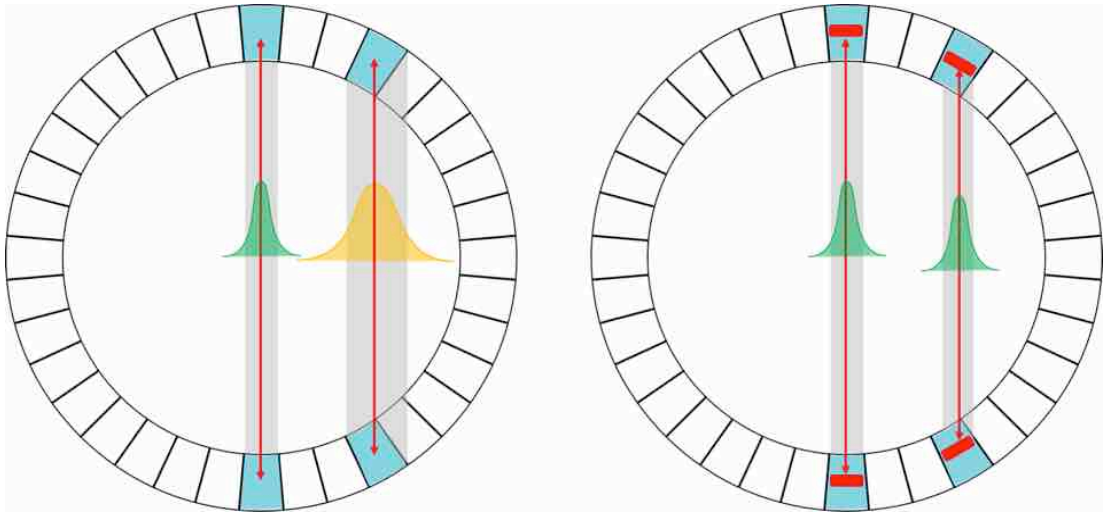

Figure 20. Impact of the Depth Of Interaction (DOI) on the blurring caused by the parallax error. Normally the two-gamma emission occurs from an off-center extended source, and the detector size widens the LOR (left). A determination of the DOI event by event prevents such a widening that would worsen the spatial resolution (right).

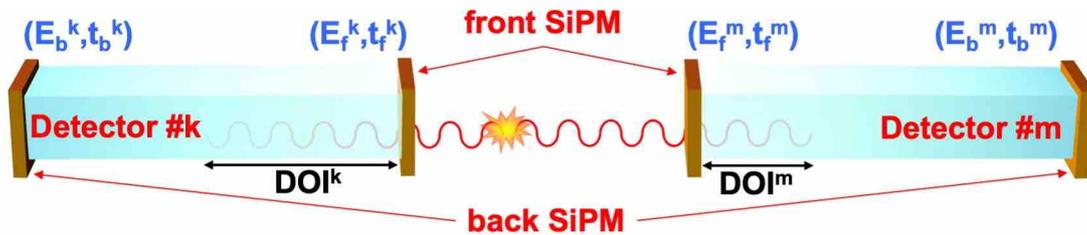

Figure 21. Scheme of the detection system employed in the TOPEM project [20]. See the text for the meaning of the indicated quantities.

An experimental verification of Eqs.3 and 4 can be done by means of a suitable lead collimator and a $^{22}$Na beta+ laboratory source, as explained in ref. [35]. The collimator makes it possible to send 511 keV gamma rays, coming from the annihilation of positrons inside the source itself, onto the LYSO scintillator with a sub-millimeter precision (Figure 22 left), thus producing an interaction at a predefined depth. The plot of Figure 22 right, produced by means of Eq.3, perfectly reproduces the tested impact positions at 1.5, 3, 5, 7, 8.5 mm. By using the same collimator one can also prove the correctness of Eq.4. Indeed, Figure 23 top shows the 2D histogram of $E_f$, $E_b$ when irradiating in three positions, whereas Figure 23 bottom shows the case of total irradiation. The red contour curve indicates the hyperbola-like locus of the 511 keV full energy peak of the detected gamma rays.

Many research groups worldwide are currently working to finalize the benefits of high resolution TOF-PET, with very promising results which will soon be available for real life clinical applications.

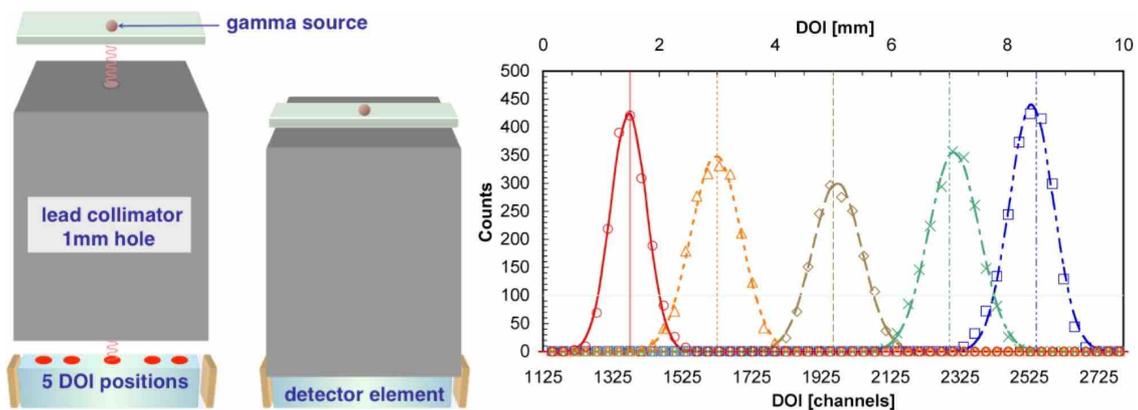

Figure 22. Left: experimental setup for the calibration of the DOI. Right: number of counts as a function of DOI, obtained by means of Eq.4, at 1.5, 3, 5, 7, 8.5 mm impact position. The primary (lower) X axis is in arbitrary units, the secondary (upper) axis is calibrated.

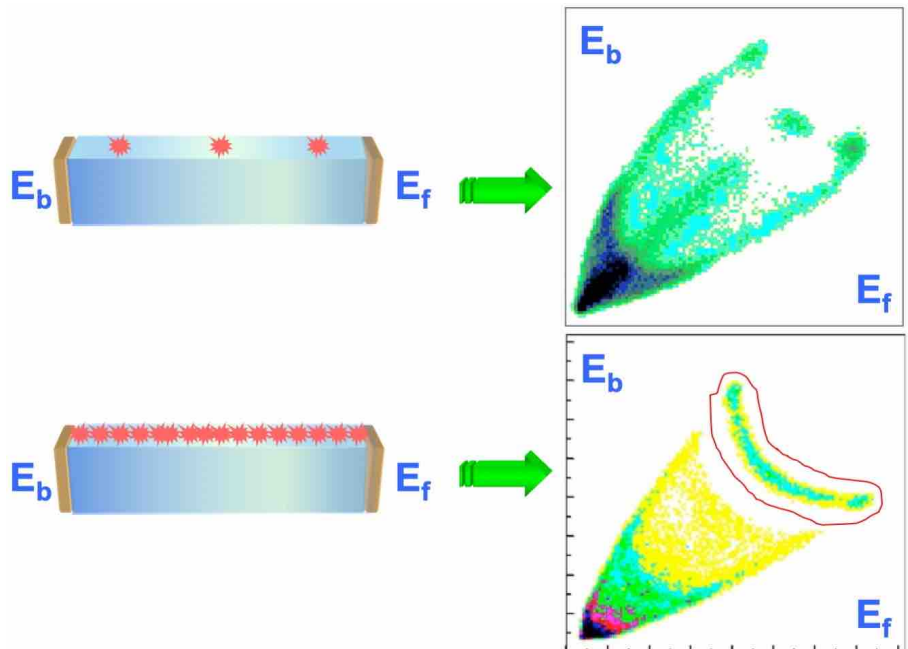

Figure 23. (Top) 2D histogram of $E_f$, $E_b$ when irradiating in three positions. Bottom: same plot in case of total irradiation. The red contour curve indicates the hyperbola-like locus of the 511 keV full energy peak of the detected gamma rays.

## 4 Applications: online monitoring of radioactive waste

When dealing with nuclear material, an important concept is the one of the so-called 3S's: safety, security, and safeguards. Safety is the protection of people from exposure to dangerous ionizing radiation and it is regulated according to several norms from each national authority. Security deals with the protection of nuclear material and facilities against malevolent acts, and the Convention on the Physical Protection of Nuclear Material was adopted in the frame of the International Atomic Energy Agency (IAEA) on 26 October 1979 in Vienna, Austria. Throughout the years about 160 countries have signed the convention. In November 2017 IAEA organized in Vienna the International Conference on Physical Protection of Nuclear Material and Nuclear Facilities, and this testifies the increasing relevance of the subject in light of the evolving worldwide situation in terms of criminality and terrorism. Last but not least, safeguards ensure that nuclear material and technology are used only for peaceful purposes and not for the development of nuclear weapons. The Non-Proliferation Treaty (NPT) was also developed in the frame of the IAEA and entered into force in 1970. As of 2019, 190 Member States are part of the NPT.

The protection of nuclear material and nuclear facilities involves storage, handling, transport, inspection, check, with both security and safety issues. Moreover, some paperwork of old packages can be lost or nonexisting (e.g. legacy waste), and people working in nuclear installations retire thus causing some possible loss of know-how: therefore maintaining the continuity of knowledge is of paramount importance. Due to the long half-life of many radioactive nuclear species, nuclear materials can last even up to hundreds of thousands years and their so far envisioned final destination is the geological repository, that is a deep underground excavation with suitable physico-chemical and mineralogical features. There are still doubts on the effective suitability and reliability of this solution, nonetheless in most countries there are ongoing activities at different stages of development.

However, even though such repositories are foreseen to last virtually forever, before closing and sealing them there will be a lot of preclosure interim activities which can last 10-100 years. During such time span there will be the need to prevent, detect and respond to accidents, theft, sabotage, unauthorized access and illegal transfer or other malicious acts. One of the tasks required to mitigate the possible problems arising when dealing with nuclear materials is monitoring, in order to check the package integrity at each stage of its life until its final deployment underground. So far monitoring has been done with conventional methods, i.e. using a number of fixed ambient radioactivity monitor stations and periodic or non-scheduled checks by operators carrying portable detection systems.

In February 2014 at the Waste Isolation Power Plant (WIPP) in New Mexico, USA, a radioactive waste drum exploded because of a previous packing mistake [36]. The event was detected with a non-specified delay by an underground ambient monitoring station, and six weeks later 21 individuals were identified through bioassay to have initially tested positive for low level amounts of internal contamination. An open question is if a constant individual drum monitoring could have prevented or mitigated the accident, by providing an early warning before the explosion when the pressure build-up would have started some local leak from the drum lid.

Nowadays new technologies could supplement if not replace the conventional monitoring methods, while providing a complete and fully digital characterization and assessment procedure for the nuclear material, in particular for radioactive waste. This is, for instance, the objective of the MICADO project recently funded by Euratom [37]. An individual and continuous monitoring should make possible to have a complete and detailed record of the history of each cask, with the goal of improving safety, security and transparency, thus minimizing the direct human intervention, accidents, mistakes, the possibility of malicious acts. Newly developed compact low-cost radiation detectors, to be directly applied to the radwaste drums, could monitor nuclear materials in place and/or during transportation, as sketched in Figure 24, thus detecting possible diversion and preventing illicit trafficking as pointed out in the DMNR project [38],[39],[40].

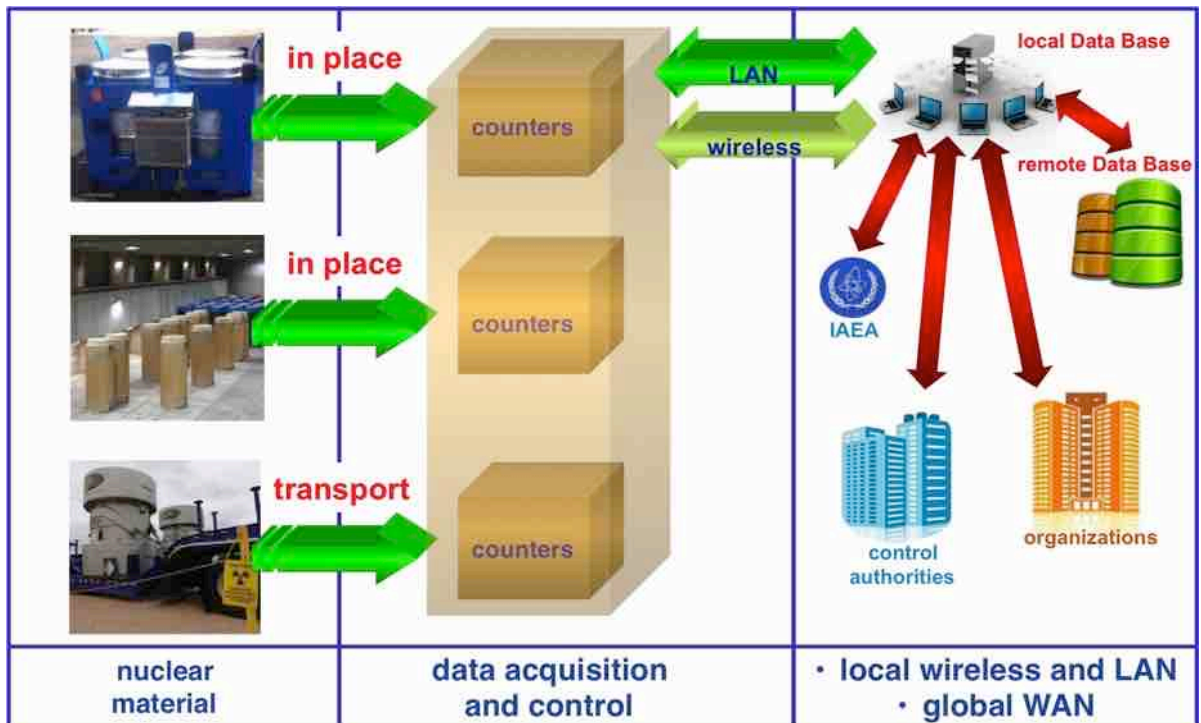

Figure 24. Scheme of an online real-time radwaste monitoring system, with connection to a redundant data base accessible to the authorized organizations. Top: low and intermediate level radwaste, mainly gamma emission. Middle: high level radwaste, mainly spent fuel emitting gamma and neutrons. Bottom: nuclear material transport monitoring.

In the following sections 4.1 and 4.2 two low-cost gamma and neutron detection technologies, suitable for being employed in a granular monitoring system, are described. Neutrons and gamma rays, due to their high penetration properties, convey information from the inside. An unexpected change in counting rate would be an indication of anomaly, and this is the reason why a low-cost compact monitoring solution should primarily focus on neutrons and gamma rays detection [41]. Figure 25 represents a very schematic summary of the nuclear fuel and radwaste cycle, in which basically all phases where gamma ray and/or neutron continuous monitoring can be useful are highlighted with appropriate symbols for transport or in-place monitoring.

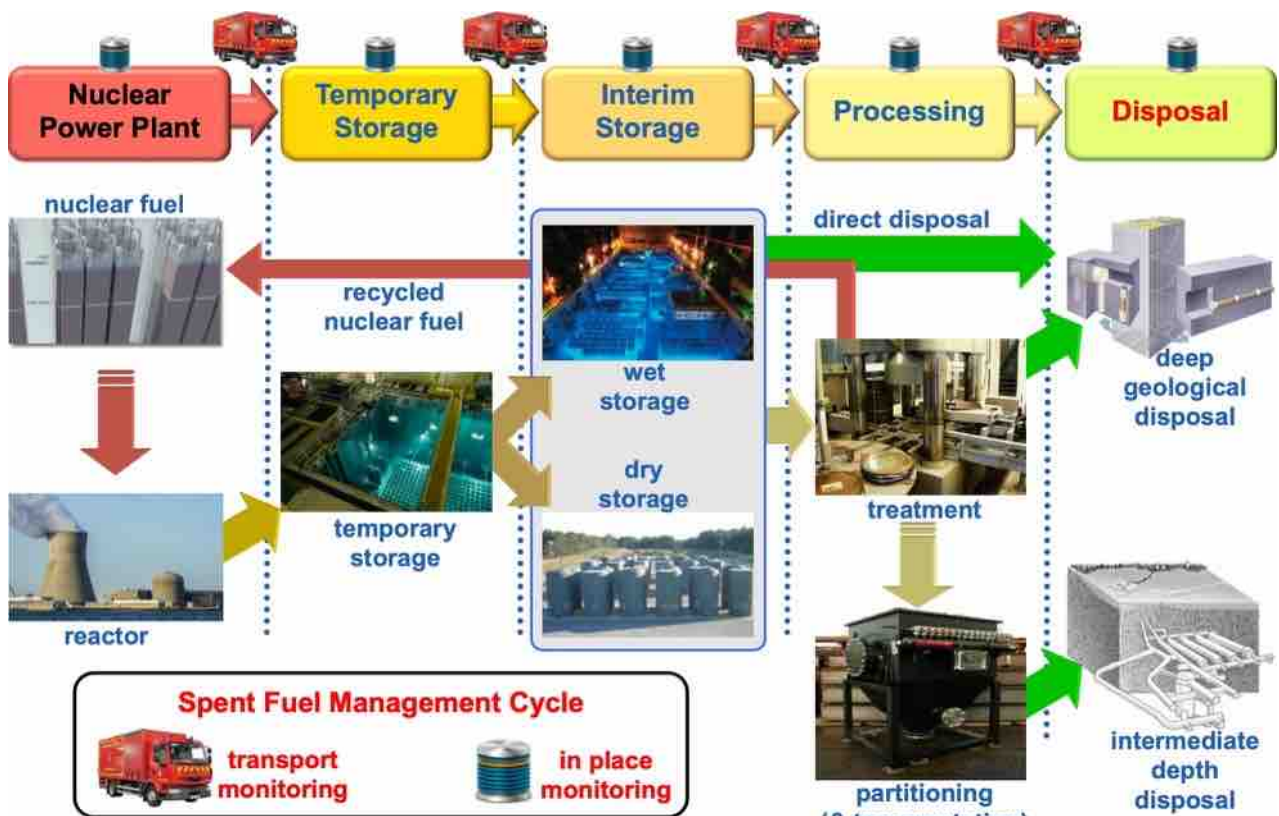

Figure 25. Schematic summary of the nuclear fuel and radwaste cycle. Basically all phases where gamma ray and/or neutron continuous monitoring can be useful are highlighted with the truck or drum symbols for transport or in-place monitoring.

### 4.1 The SciFi linear gamma ray counter

The operational principle of the SciFi gamma ray counter, schematically illustrated in Figure 26, is based on a plastic scintillating optical fiber read out at each end by a SiPM. Whenever the gamma radiation interacts with the fiber a very short light flash is produced, it propagates in both directions along the fiber and can be detected by the two SiPMs. Imposing the simultaneous detection at both ends strongly suppresses the spurious noise. Such a detector, whose features and limitations are described in the following, can be assembled either in form of a rigid pipe, made from plastic or aluminum, or as a flexible rubber tube [42].

The scintillating fiber, already introduced in section 2.3 consists of a scintillating core of polystyrene surrounded by a thin cladding made from PMMA. The light propagates inside the fiber by total reflection at the core-cladding interface, virtually with no disturbance caused by the external medium (mainly air but also materials coming in contact with the fiber). Table 1 lists the main features of the 1 mm diameter BCF20 fiber produced by Saint Gobain [14]. The gamma ray interaction with the fiber occurs almost exclusively by means of Compton scattering [43], i.e. the gamma ray hits an electron in the fiber and is scattered away. The electron moves into the fiber, in a random direction, and releases (part of) its kinetic energy which is converted into scintillation light emitted isotropically.

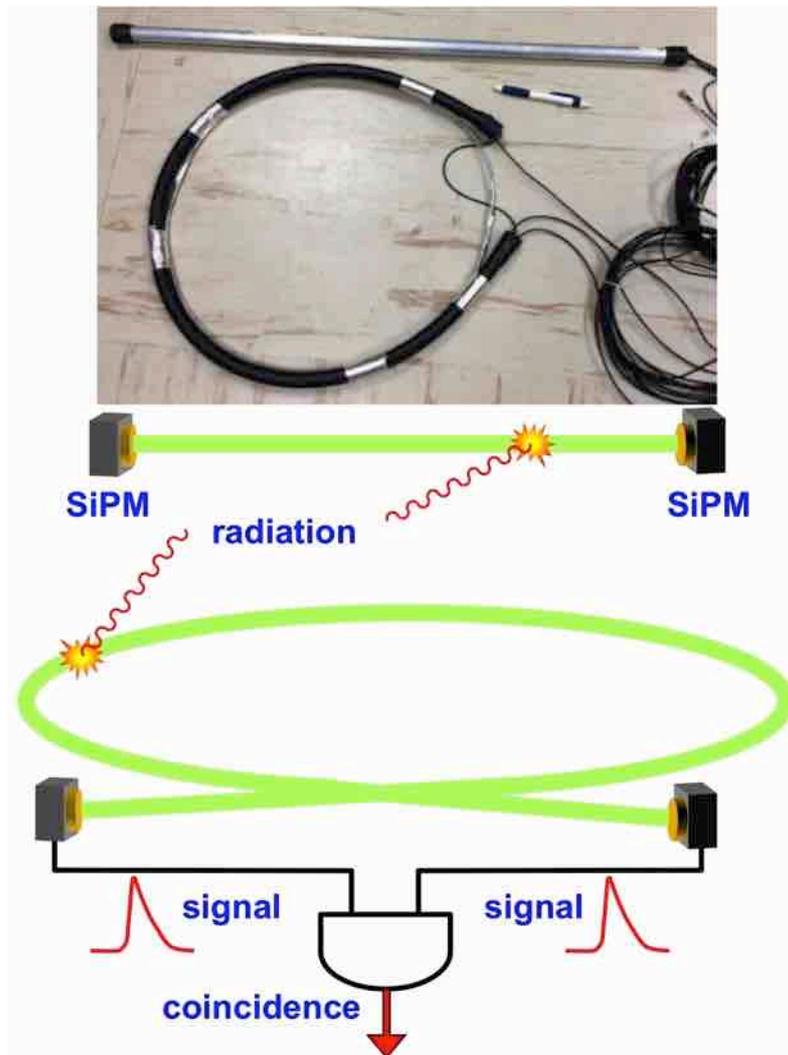

Figure 26. The operational principle of the SciFi gamma ray counter, based on a plastic scintillating optical fiber read out at each end by a SiPM. Whenever the gamma radiation interacts with the fiber a very short light flash is produced which propagates in both directions along the fiber and can be detected by the two SiPMs.

Table 1. Properties of the 1mm diameter BCF20 green plastic scintillating fibers [14].

| Core material | Polystyrene |
|---|---|
| Core refractive index | 1.60 |
| Density | 1.05 |
| Cladding material | Acrylic |
| Cladding refractive index | 1.49 |
| Cladding thickness | 3% of fiber diameter |
| Numerical aperture | 0.58 |
| Trapping efficiency | ≈ 6% |
| Emission color | green |
| Emission peak, nm | ≈ 500 |
| Decay time, ns | ≈ 3 |
| 1/e length m | >3.5 |
| n. of photons per MeV | ≈ 8000 |

Figure 27 illustrates the geometry and the optical working principle of the fiber, with the total reflection angle determined by the ratio between the refractive indexes of cladding and core. The trapping efficiency, that is the fraction of light transported by total reflection, can be calculated analytically for trajectories originated on the simmetry axis and is about 3.5% in both directions. Numerical simulations taking into account also the skew rays provide the effective value of about 6%. Not all of the trapped light is transported to the fiber ends, because of some attenuation that occurs with an exponential law governed by the fiber attenuation length, a parameter indicating the distance where the light intensity reduces to a fraction *1/e*. Figure 28 shows the result of numerical simulations of the energy released in the fiber by (the energetic electrons produced by) gamma rays [42]. One can immediately see that the average deposited energy is rather constant around 180 keV for primary gamma rays above 500 keV, decreasing to about 80 keV at 100 keV primary gamma energy.

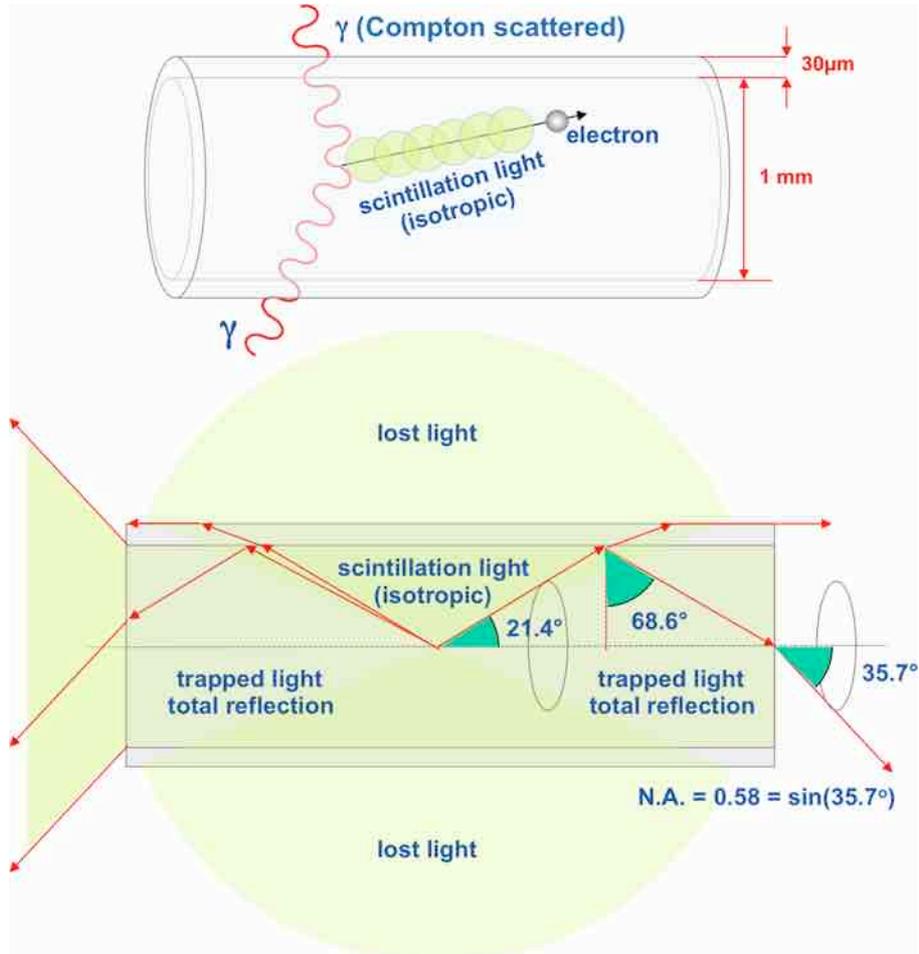

Figure 27. Geometry and optical working principle of the fiber. The total reflection angle is determined by the ratio between the refractive indexes of cladding and core. The trapping efficiency, that is the fraction of light transported by total reflection, can be calculated analytically for trajectories originated on the simmetry axis and is about 3.5% in both directions. Numerical simulations taking into account also the skew rays provide the effective value of about 6%.

The SiPM photodetector [21]-[32], already introduced in section 3.2, is an integrated array of silicon photodiodes operating in Geiger mode with a common output. This means that each photodiode, or cell, is reverse biased slightly above its breakdown voltage where it keeps stable for quite some time (even ≈ 10-100 milliseconds) during which it is sensitive to visible photons. Whenever a visible photon interacts it takes an *e-h* pair into the conduction band, and the electron and the hole start moving toward the opposite polarity electrodes. Due to the overvoltage the electron, whose mobility is larger, can get enough kinetic energy to hit and free other electrons thus generating an avalanche (Figure 29). Each cell is intrinsically protected by an integrated quenching resistor that prevents the physical breakdown by momentarily lowering the effective voltage and stopping the avalanche multiplication (Figure 30). The electrical gain is typically of the order of $10^6$, i.e. one photon frees one electron that the avalanche multiplies to $10^6$ electrons. The output signal is quasi-digital, as it is the superposition of a number of identical signals, one for each cell fired (by a photon or noise). The array constituting the SiPM is typically made of thousands or tens of thousands cells, depending on the device size. As of now on the market there are SiPMs from 1mm × 1mm up to 6mm × 6mm, and also arrays of SiPMs.

Each SiPM cell is sensitive to single photons but unfortunately it can also be randomly triggered by thermal reasons in the semiconductor: as said before, it is stable over the breakdown voltage for some time, then it randomly triggers an avalanche. Such an avalanche is indistinguishable from a real photon detection and is considered as 1-cell thermal noise. Due to statistical reasons, the probability of two or more cells simultaneously triggered by thermal noise is lower and lower. In order to suppress spurious events, not originated by a real gamma interaction in the scintillating fiber, in addition to the coincidence between the two SiPMs an output signal of at least four cells is required in each SiPM, basically excluding the very low probability event of the random simultaneous thermal signal generation in 4+4 cells.

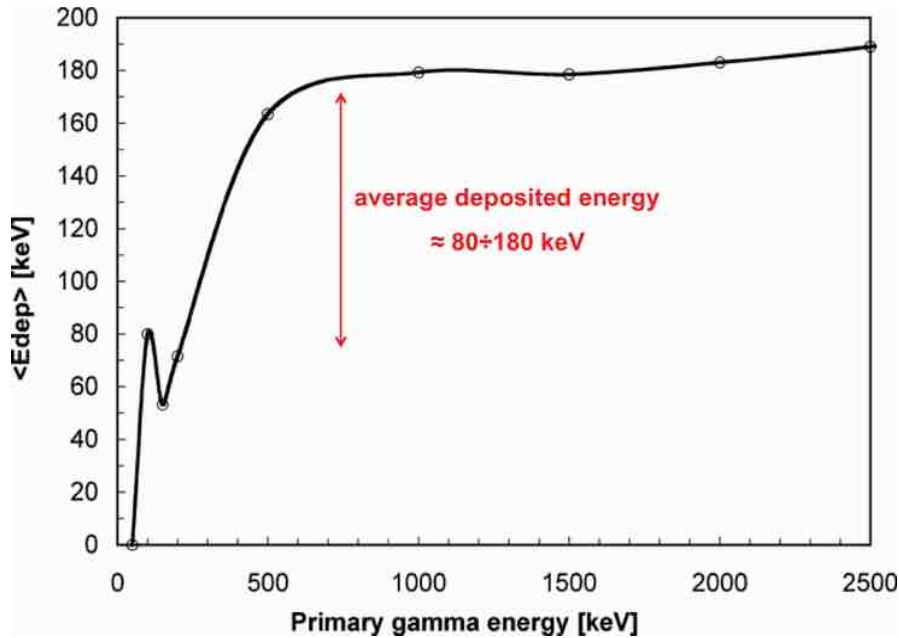

Figure 28. Numerical simulation of the energy released in the fiber by (the energetic electrons produced by) gamma rays.

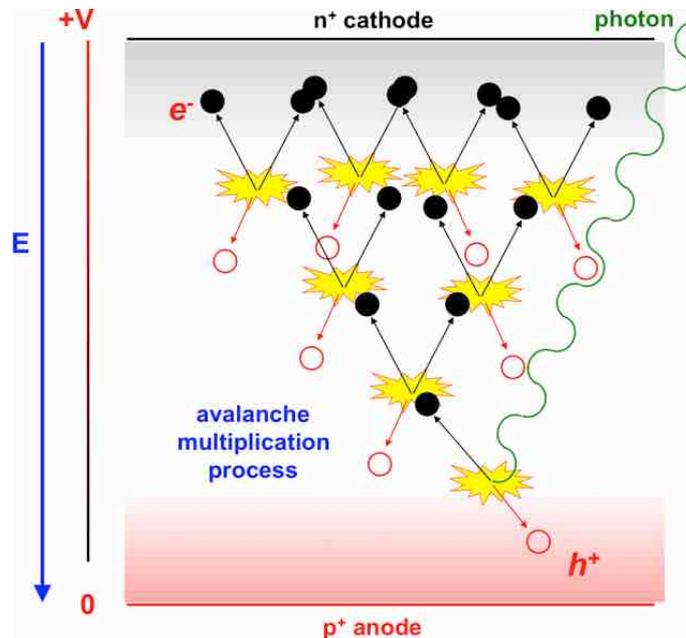

Figure 29. Pictorial representation of a visible photon interacting in the depletion region of a SiPM cell. The photon is absorbed and frees an *e-h* pair, the electron (full circles) and the hole (open circles) start moving toward the opposite polarity electrodes. Due to the overvoltage the electron, whose mobility is larger, can get enough kinetic energy to hit and free other electrons thus generating an avalanche.

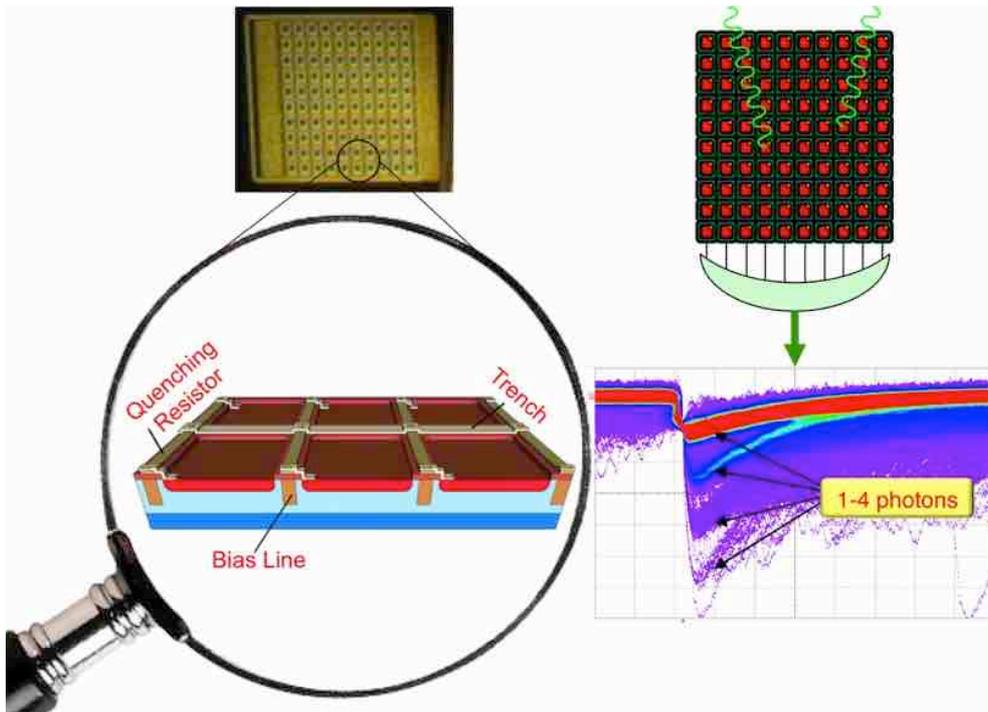

Figure 30. Left: microphotograph of a 100-cell 1mm × 1mm SiPM. Each cell is intrinsically protected by an integrated quenching resistor that prevents the physical breakdown by momentarily lowering the effective voltage and stopping the avalanche multiplication. Also present are trenches, filled by opaque materials, optically isolating each cell in order to prevent cross-talk. Right: the output signal is quasi-digital, as it is the sum of a number of identical signals, one for each fired cell. This is evident in the oscilloscope snapshot showing discrete loci for one to four cell signals.

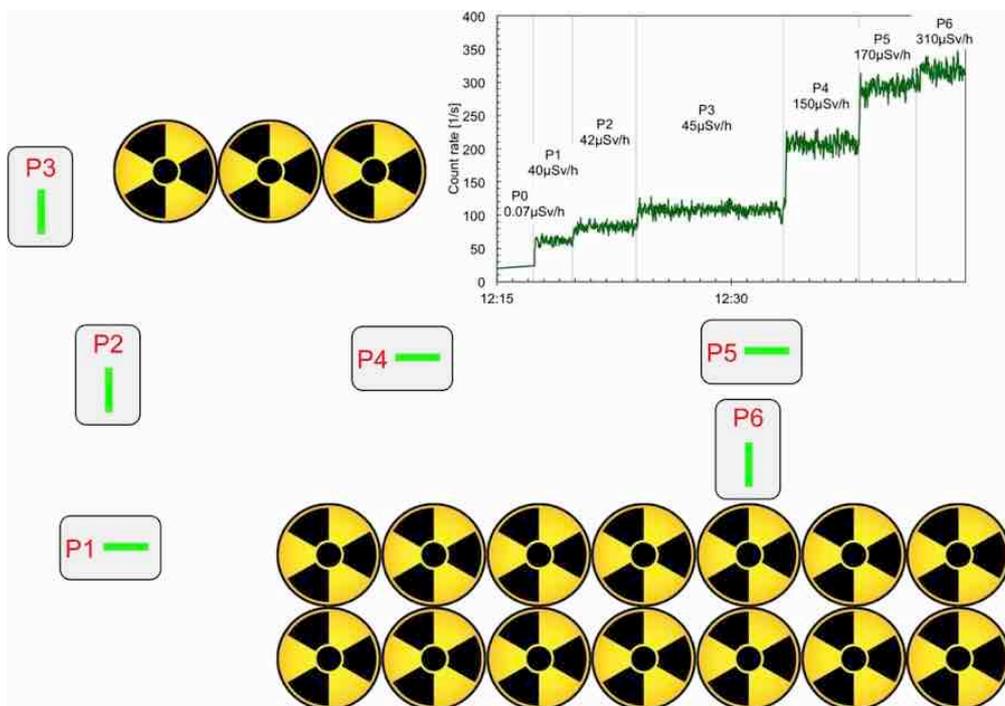

Figure 31. Sketch of a test performed with ILW in a real storage site. The SciFi sensor was installed on a cart and placed at six positions with increasing dose rate. The plot shows the correspondence between the dose rate measured by a reference Geiger counter and the SciFi.

SciFi counters have been extensively tested with cosmic rays and laboratory sources, then at radwaste storage sites with drums containing Low Level Waste (LLW), Intermediate Level Waste (ILW) and also High Level Waste (HLW, spent fuel). Despite their low detection efficiency around 1%, the results have been more than encouraging and their use is now foreseen in the framework of the already mentioned MICADO project.

Figure 31 is the sketch of a test performed with ILW in a real storage site. The SciFi sensor was installed on a cart and placed at six positions with increasing dose rate. The plot shows the correspondence between the dose rate measured by a reference Geiger counter and the SciFi.

*4.2   The SiLiF neutron counter*

The detection of neutrons is required in many fundamental and applied science fields, as well as in several industrial applications. Moreover, and more important, it is mandatory for nuclear safety and security reasons. International organizations for nuclear safeguards (e.g. IAEA) rely heavily on neutron detection techniques for the verification of declared nuclear materials and for monitoring purposes. The centerpiece in neutron detection is $^3$He, an artificially produced gas which undergoes the $^3$He(n,p)$^3$H reaction with thermal neutrons with a high cross section (5300 b) [44]. A suitable gas detector filled with $^3$He features a high neutron detection efficiency and a good gamma-ray insensitivity (I remark that in general wherever there are neutrons there are also gamma rays).

While $^3$He was readily available, because of the reliability, safety, ease of use, gamma-ray insensitivity, and high intrinsic thermal neutron detection efficiency of $^3$He-based detectors, there was no real need for alternative detector technologies. During the last two decades the decline of the $^3$He gas supply, as well as its increasing price, has triggered international efforts to develop neutron detectors that make use of alternative materials and/or techniques.

The direct detection of neutrons is not possible, as it is a neutral particle thus not directly ionizing matter. What is needed is a so called converter, capable of converting the incoming neutrons into charged particles to be detected by means of suitable detectors. Indeed the candidate materials are not many: apart from $^3$He one can list just a few isotopes, with their thermal neutron reaction cross section indicated in parenthesis, $^6$Li (940 b), $^{10}$B (3800 b), $^{113}$Cd (20 kb), $^{155}$Gd (61 kb) and $^{157}$Gd (254 kb). The neutron absorption cross section at low energy is basically proportional to the inverse of the neutron velocity. Despite their very high cross section, upon neutron capture Cadmium and Gadolinium decay by emitting a large number of energetic gamma rays. This would make quite difficult disentangling background gamma rays from those produced by the neutron interaction in the converter. Therefore the only converters deemed useful are $^6$Li and $^{10}$B, as they decay by particle emission. Indeed $^{235}$U (600 b fission cross section) could also be useful, but it is a strategic radioactive material and its use is strongly restricted by laws and international treaties.

Upon neutron capture $^6$Li and $^{10}$B decay according to the following reactions:

$$^6Li + n \rightarrow\ ^3H\ (2.73\ MeV) + \alpha\ (2.05\ MeV)$$
$$^{10}B + n \rightarrow\ ^7Li\ (1.01\ MeV) + \alpha\ (1.78\ MeV)\ \text{ground state, BR} \approx 6\%$$
$$^{10}B + n \rightarrow\ ^7Li\ (0.84\ MeV) + \alpha\ (1.47\ MeV) + \gamma\ (0.48\ MeV)\ \text{excited state, BR} \approx 94\%$$

The produced charged particles lose some of their energy passing through the converter, the amount of loss depending on the converter material and its thickness, the particle type and its energy. If deciding for a solid state detection technique the choice of film thickness that can be practically used is limited, otherwise the choice should be toward a gas-based detector. There are pros and cons in both solutions, and this section will deal with a solid state detection technique where a converter layer is placed in front of a silicon detector in order to detect one of the reaction products as a signature of the neutron interaction.

The reaction products entering the detector must have sufficient energy in order to produce signals large enough to discriminate charged particles from the gamma rays. By comparing the above listed reactions one immediately realizes that the triton and the alpha produced by neutron interaction with $^6$Li have higher kinetic energy than the alpha and $^7$Li produced by $^{10}$B. Moreover, since the lighter the particle the longer its range in matter, the triton and alpha from $^6$Li are less hindered by dead layers and consequently are better suited for detection with semiconductor detectors, even though the neutron cross section in $^{10}$B is higher than in $^6$Li. Therefore $^6$Li was chosen as converter nuclear species, but because it is chemically highly reactive it was decided to use it as LiF compound, enriched at 95% in $^6$Li, that is a stable salt moderately inexpensive, hence the SiLiF name [45]-[52].

A thin layer of $^6$LiF converter (1-16 µm depending on the required performance) is deposited on a substrate. Two such converters are placed on either sides of a fully depleted silicon detector, i.e. that is capable to detect particles entering from both faces. The resulting sandwich is the so called SiLiF neutron detector. It is mainly sensitive to thermal neutrons, because of the $1/v_n$ decrease of the cross section and then, in order to increase the sensitivity to neutrons of higher energy, one usually surrounds the detector with a light material capable of moderating (i.e. slowing down) the neutrons hence called moderator. A sketch of the nuclear reaction of interest and of the detector setup is depicted in Figure 32.

The SiLiF technique has been extensively tested with Americium-Beryllium (AmBe) neutron sources, which also emit copious amounts of gamma rays, moderated by polyethylene. The (in)sensitivity to gamma rays was tested with laboratory sources of well known activity. An example of the spectrum of the energy deposited into the silicon detector is shown in Figure 33, where three different tests done with converters 1.6 μm thick are highlighted in different colors: (1) measurement of the background with the fully equipped detector; (2) measurement of the impact of the AmBe source on the silicon, by removing the two converters; (3) final measurement with the AmBe source and the SiLiF fully equipped. In the test n.3 one can immediately see the two bumps corresponding to the detection of the triton and of the alpha. Choosing a suitable threshold one decides which counts are to be attributed to neutrons and which are rejected, the higher the threshold the lower the neutron detection efficiency and the better the gamma/neutron discrimination. A typical measured gamma/neutron rejection factor, with the threshold set at 1.5 MeV, is $10^{-7}$.

Tests of SiLiF detectors were performed at neutron beam facilities (n_TOF at CERN [54] and ISIS at Rutherford Appleton Laboratory [55]) and are currently in use. A successful test was also done with spent fuel casks in a radwaste storage facility, thus proving that these detectors can be exploited for nuclear material monitoring. The SiLiF technique provides a valid alternative to $^3$He detectors useful for several applications, with valuable benefits in terms of cost, low-voltage operation, compactness, robustness, manageability, detection efficiency and gamma/neutron discrimination.

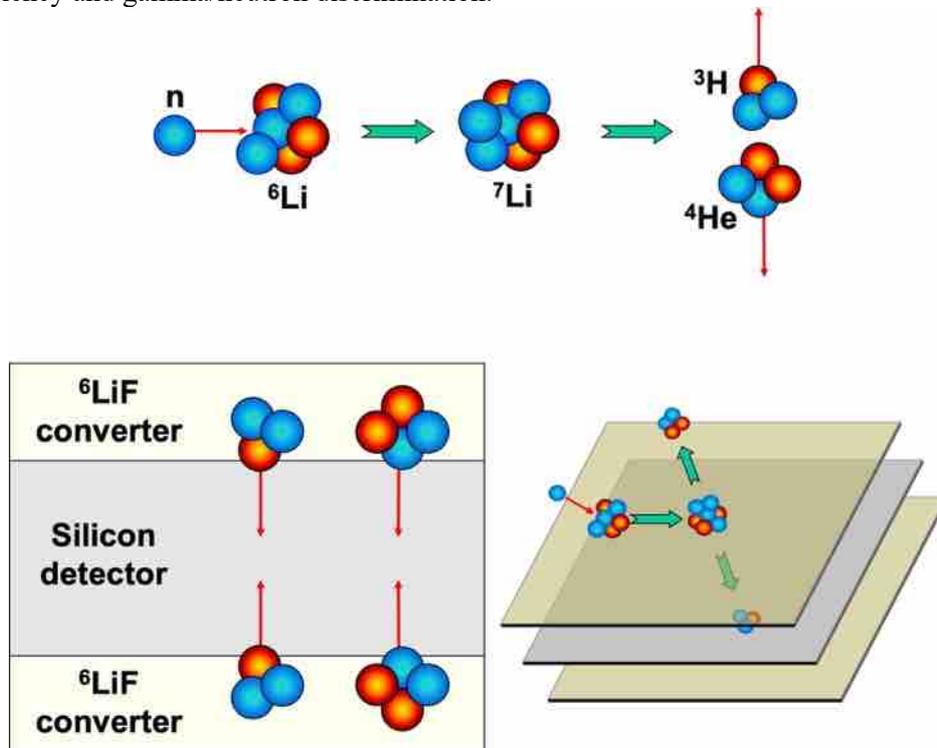

Figure 32. Top: sketch of the $^6$Li(n,t)α neutron converting nuclear reaction. Bottom: scheme of the SiLiF detector setup.

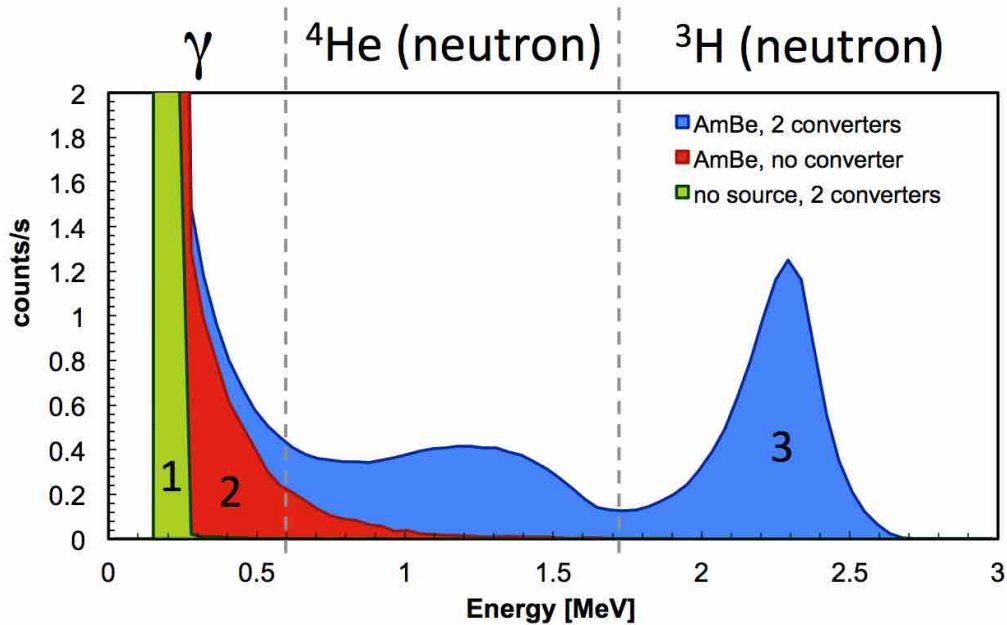

Figure 33. Spectrum of the energy deposited into the silicon detector. Three different tests were done, the converter layer thickness was 1.6 μm. (1) measurement of the background with the fully equipped detector; (2) measurement of the impact of the AmBe source on the silicon, by removing the two converters; (3) measurement with the AmBe source and the SiLiF fully equipped. In the test n.3 the two bumps corresponding to the detection of the triton and of the alpha are clearly visible.

# 5  Conclusion

In this paper a short ride through a few examples of applications of nuclear detection techniques has shown how ideas, methods and devices originally meant for fundamental research can migrate to neighboring fields, like the diagnostics of accelerated ion beams, or to social and industrial applications like paramount medical diagnostics and nuclear material monitoring. This is maily due to two peculiarities offered by nuclear physics: the capability to investigate the inside of things because of the high penetration of radiations like gamma rays and neutrons, and the capability to be sensitive to single nuclei with their properties, interactions and decays. A lot of other applications were not mentioned here, and many more new ones will certainly be invented and become available in the forthcoming years.

# 6  Acknowledgments

The work in this paper was partially done within the framework of European Union's Horizon 2020 research and innovation programme under grant agreement No 847641, project MICADO (Measurement and Instrumentation for Cleaning And Decommissioning Operations).

I am grateful to Angela Bonaccorso, Ignazio Bombaci, Maria Agnese Ciocci and Valeria Rosso of INFN Pisa, for inviting me to the Pisa Summer School 2019, thus giving me the opportunity to share my views with several people, mostly students, that I enjoy enormously.